\documentclass[conference]{IEEEtran}
\IEEEoverridecommandlockouts

\def\BibTeX{{\rm B\kern-.05em{\sc i\kern-.025em b}\kern-.08em
    T\kern-.1667em\lower.7ex\hbox{E}\kern-.125emX}}

\usepackage[nolist]{acronym}
\begin{acronym}
    \acro{OSS}{open source software}
    \acro{NLP}{natural language processing}
    \acro{NLU}{natural language understanding}
    \acro{IR}{information retrieval}
    \acro{ML}{machine learning}
    \acro{LLM}{large language model}
    \acro{LDA}{latent Dirichlet allocation}
    \acro{SVM}{support vector machine}
    \acro{NBC}{naive Bayes}
    \acro{VSM}{vector space model}
    \acro{LSI}{latent semantic indexing}
    \acro{JS}{probabilistic Jensen and Shannon model}
    \acro{CNN}{convolutional neural network}
    \acro{d2v}{doc2vec}
    \acro{SCDV}{sparse composite document vector}
    \acro{ELMo}{Embeddings from Language Models}
    \acro{GAT}{graph attention network}
    \acro{BERT}{Bidirectional Encoder Representations from Transformers}
    \acro{ULMFiT}{Universal Language Model Fine-tuning for Text Classification}
    \acro{LSTM}{long short-term memory}
    \acro{NFR}{non-functional requirements}
    \acro{F}{functional requirements}
    \acro{OS}{oversampling}
    \acro{US}{undersampling}
    \acro{ES}{early stopping}
    \acro{TL}{trace link}
    \acro{TLR}{traceability link recovery}
    \acro{WMD}{Word Mover's Distance}
    \acro{PNFR}{PROMISE NFR}
    \acro{NoRBERT}{Non-functional and functional Requirements classification using BERT}
    \acro{TLearn}{transfer learning}
    \acro{RE}{requirements engineering}
    \acro{FTLR}{Fine-grained Traceability Link Recovery}
    \acro{MAP}{mean average precision}
    \acro{AP}{average precision}
    \acro{RNN}{recurrent neural network}
    \acro{FFNN}{feedforward neural network}
    \acro{RAG}{retrieval-augmented generation}
    \acro{CoT}{chain-of-thought}
\end{acronym}

\usepackage{hyperref}
\hypersetup{
	final,
	hidelinks,
}

\usepackage{amsmath}
\usepackage{amssymb}
\usepackage{nicefrac}

\usepackage[nolist]{acronym}
\usepackage{xspace}

\usepackage[table]{xcolor}
\usepackage{array}
\usepackage{tabularx}
\newcolumntype{Y}{>{\raggedleft\arraybackslash}X}
\newcolumntype{Z}{>{\centering\let\newline\\\arraybackslash\hspace{0pt}}X}
\newcolumntype{H}{>{\setbox0=\hbox\bgroup}c<{\egroup}@{}}
\usepackage{booktabs}
\usepackage{multirow}
\usepackage{colortbl}
\newsavebox\CBox

\usepackage{rotating}

\RequirePackage{soul}

\definecolor{KITGreen}{RGB}{0,150,130}
\definecolor{KITBlue}{RGB}{70,100,170}
\definecolor{KITRed}{RGB}{162,34,35}
\definecolor{KITOrange}{RGB}{223,155,27}
\definecolor{KITMaygreen}{RGB}{140,182,60}
\definecolor{KITYellow}{RGB}{252,229,0}
\definecolor{KITBrown}{RGB}{167,130,46}
\definecolor{KITPurple}{RGB}{163,16,124}
\definecolor{KITCyan}{RGB}{35,161,224}
\definecolor{BERTColor}{RGB}{255,192,0}

\definecolor{lightlightgray}{rgb}{.96,.96,.96}
\definecolor{russet}{rgb}{0.5, 0.27, 0.11}
\definecolor{BurntOrange}{RGB}{242,134,9}

\definecolor{kit-green}{RGB}{0, 150, 130}
\definecolor{kit-green100}{RGB}{0, 150, 130}
\definecolor{kit-green90}{rgb}{0.1, 0.6294, 0.5588}
\definecolor{kit-green80}{rgb}{0.2, 0.6706, 0.6078}
\definecolor{kit-green75}{rgb}{0.25, 0.6912, 0.6324}
\definecolor{kit-green70}{rgb}{0.3, 0.7118, 0.6569}
\definecolor{kit-green60}{rgb}{0.4, 0.7529, 0.7059}
\definecolor{kit-green50}{rgb}{0.5, 0.7941, 0.7549}
\definecolor{kit-green40}{rgb}{0.6, 0.8353, 0.8039}
\definecolor{kit-green30}{rgb}{0.7, 0.8765, 0.8529}
\definecolor{kit-green25}{rgb}{0.75, 0.8971, 0.8775}
\definecolor{kit-green20}{rgb}{0.8, 0.9176, 0.902}
\definecolor{kit-green15}{rgb}{0.85, 0.9382, 0.9265}
\definecolor{kit-green10}{rgb}{0.9, 0.9588, 0.951}
\definecolor{kit-green5}{rgb}{0.95, 0.9794, 0.9755}

\definecolor{kit-blue}{RGB}{70, 100, 170}
\definecolor{kit-blue100}{RGB}{70, 100, 170}
\definecolor{kit-blue90}{rgb}{0.3471, 0.4529, 0.7}
\definecolor{kit-blue80}{rgb}{0.4196, 0.5137, 0.7333}
\definecolor{kit-blue75}{rgb}{0.4559, 0.5441, 0.75}
\definecolor{kit-blue70}{rgb}{0.4922, 0.5745, 0.7667}
\definecolor{kit-blue60}{rgb}{0.5647, 0.6353, 0.8}
\definecolor{kit-blue50}{rgb}{0.6373, 0.6961, 0.8333}
\definecolor{kit-blue40}{rgb}{0.7098, 0.7569, 0.8667}
\definecolor{kit-blue30}{rgb}{0.7824, 0.8176, 0.9}
\definecolor{kit-blue25}{rgb}{0.8186, 0.848, 0.9167}
\definecolor{kit-blue20}{rgb}{0.8549, 0.8784, 0.9333}
\definecolor{kit-blue15}{rgb}{0.8912, 0.9088, 0.95}
\definecolor{kit-blue10}{rgb}{0.9275, 0.9392, 0.9667}
\definecolor{kit-blue5}{rgb}{0.9637, 0.9696, 0.9833}

\definecolor{kit-red}{RGB}{162, 34, 35}
\definecolor{kit-red100}{RGB}{162, 34, 35}
\definecolor{kit-red90}{rgb}{0.6718, 0.22, 0.2235}
\definecolor{kit-red80}{rgb}{0.7082, 0.3067, 0.3098}
\definecolor{kit-red75}{rgb}{0.7265, 0.35, 0.3529}
\definecolor{kit-red70}{rgb}{0.7447, 0.3933, 0.3961}
\definecolor{kit-red60}{rgb}{0.7812, 0.48, 0.4824}
\definecolor{kit-red50}{rgb}{0.8176, 0.5667, 0.5686}
\definecolor{kit-red40}{rgb}{0.8541, 0.6533, 0.6549}
\definecolor{kit-red30}{rgb}{0.8906, 0.74, 0.7412}
\definecolor{kit-red25}{rgb}{0.9088, 0.7833, 0.7843}
\definecolor{kit-red20}{rgb}{0.9271, 0.8267, 0.8275}
\definecolor{kit-red15}{rgb}{0.9453, 0.87, 0.8706}
\definecolor{kit-red10}{rgb}{0.9635, 0.9133, 0.9137}
\definecolor{kit-red5}{rgb}{0.9818, 0.9567, 0.9569}

\definecolor{kit-yellow}{RGB}{252, 229, 0}
\definecolor{kit-yellow100}{RGB}{252, 229, 0}
\definecolor{kit-yellow90}{rgb}{0.9894, 0.9082, 0.1}
\definecolor{kit-yellow80}{rgb}{0.9906, 0.9184, 0.2}
\definecolor{kit-yellow75}{rgb}{0.9912, 0.9235, 0.25}
\definecolor{kit-yellow70}{rgb}{0.9918, 0.9286, 0.3}
\definecolor{kit-yellow60}{rgb}{0.9929, 0.9388, 0.4}
\definecolor{kit-yellow50}{rgb}{0.9941, 0.949, 0.5}
\definecolor{kit-yellow40}{rgb}{0.9953, 0.9592, 0.6}
\definecolor{kit-yellow30}{rgb}{0.9965, 0.9694, 0.7}
\definecolor{kit-yellow25}{rgb}{0.9971, 0.9745, 0.75}
\definecolor{kit-yellow20}{rgb}{0.9976, 0.9796, 0.8}
\definecolor{kit-yellow15}{rgb}{0.9982, 0.9847, 0.85}
\definecolor{kit-yellow10}{rgb}{0.9988, 0.9898, 0.9}
\definecolor{kit-yellow5}{rgb}{0.9994, 0.9949, 0.95}

\definecolor{kit-orange}{RGB}{223, 155, 27}
\definecolor{kit-orange100}{RGB}{223, 155, 27}
\definecolor{kit-orange90}{rgb}{0.8871, 0.6471, 0.1953}
\definecolor{kit-orange80}{rgb}{0.8996, 0.6863, 0.2847}
\definecolor{kit-orange75}{rgb}{0.9059, 0.7059, 0.3294}
\definecolor{kit-orange70}{rgb}{0.9122, 0.7255, 0.3741}
\definecolor{kit-orange60}{rgb}{0.9247, 0.7647, 0.4635}
\definecolor{kit-orange50}{rgb}{0.9373, 0.8039, 0.5529}
\definecolor{kit-orange40}{rgb}{0.9498, 0.8431, 0.6424}
\definecolor{kit-orange30}{rgb}{0.9624, 0.8824, 0.7318}
\definecolor{kit-orange25}{rgb}{0.9686, 0.902, 0.7765}
\definecolor{kit-orange20}{rgb}{0.9749, 0.9216, 0.8212}
\definecolor{kit-orange15}{rgb}{0.9812, 0.9412, 0.8659}
\definecolor{kit-orange10}{rgb}{0.9875, 0.9608, 0.9106}
\definecolor{kit-orange5}{rgb}{0.9937, 0.9804, 0.9553}

\definecolor{kit-lightgreen}{RGB}{140, 182, 60}
\definecolor{kit-lightgreen100}{RGB}{140, 182, 60}
\definecolor{kit-lightgreen90}{rgb}{0.5941, 0.7424, 0.3118}
\definecolor{kit-lightgreen80}{rgb}{0.6392, 0.771, 0.3882}
\definecolor{kit-lightgreen75}{rgb}{0.6618, 0.7853, 0.4265}
\definecolor{kit-lightgreen70}{rgb}{0.6843, 0.7996, 0.4647}
\definecolor{kit-lightgreen60}{rgb}{0.7294, 0.8282, 0.5412}
\definecolor{kit-lightgreen50}{rgb}{0.7745, 0.8569, 0.6176}
\definecolor{kit-lightgreen40}{rgb}{0.8196, 0.8855, 0.6941}
\definecolor{kit-lightgreen30}{rgb}{0.8647, 0.9141, 0.7706}
\definecolor{kit-lightgreen25}{rgb}{0.8873, 0.9284, 0.8088}
\definecolor{kit-lightgreen20}{rgb}{0.9098, 0.9427, 0.8471}
\definecolor{kit-lightgreen15}{rgb}{0.9324, 0.9571, 0.8853}
\definecolor{kit-lightgreen10}{rgb}{0.9549, 0.9714, 0.9235}
\definecolor{kit-lightgreen5}{rgb}{0.9775, 0.9857, 0.9618}

\definecolor{kit-purple}{RGB}{163, 16, 124}
\definecolor{kit-purple100}{RGB}{163, 16, 124}
\definecolor{kit-purple90}{rgb}{0.6753, 0.1565, 0.5376}
\definecolor{kit-purple80}{rgb}{0.7114, 0.2502, 0.589}
\definecolor{kit-purple75}{rgb}{0.7294, 0.2971, 0.6147}
\definecolor{kit-purple70}{rgb}{0.7475, 0.3439, 0.6404}
\definecolor{kit-purple60}{rgb}{0.7835, 0.4376, 0.6918}
\definecolor{kit-purple50}{rgb}{0.8196, 0.5314, 0.7431}
\definecolor{kit-purple40}{rgb}{0.8557, 0.6251, 0.7945}
\definecolor{kit-purple30}{rgb}{0.8918, 0.7188, 0.8459}
\definecolor{kit-purple25}{rgb}{0.9098, 0.7657, 0.8716}
\definecolor{kit-purple20}{rgb}{0.9278, 0.8125, 0.8973}
\definecolor{kit-purple15}{rgb}{0.9459, 0.8594, 0.9229}
\definecolor{kit-purple10}{rgb}{0.9639, 0.9063, 0.9486}
\definecolor{kit-purple5}{rgb}{0.982, 0.9531, 0.9743}

\definecolor{kit-brown}{RGB}{167, 130, 46}
\definecolor{kit-brown100}{RGB}{167, 130, 46}
\definecolor{kit-brown90}{rgb}{0.6894, 0.5588, 0.2624}
\definecolor{kit-brown80}{rgb}{0.7239, 0.6078, 0.3443}
\definecolor{kit-brown75}{rgb}{0.7412, 0.6324, 0.3853}
\definecolor{kit-brown70}{rgb}{0.7584, 0.6569, 0.4263}
\definecolor{kit-brown60}{rgb}{0.7929, 0.7059, 0.5082}
\definecolor{kit-brown50}{rgb}{0.8275, 0.7549, 0.5902}
\definecolor{kit-brown40}{rgb}{0.862, 0.8039, 0.6722}
\definecolor{kit-brown30}{rgb}{0.8965, 0.8529, 0.7541}
\definecolor{kit-brown25}{rgb}{0.9137, 0.8775, 0.7951}
\definecolor{kit-brown20}{rgb}{0.931, 0.902, 0.8361}
\definecolor{kit-brown15}{rgb}{0.9482, 0.9265, 0.8771}
\definecolor{kit-brown10}{rgb}{0.9655, 0.951, 0.918}
\definecolor{kit-brown5}{rgb}{0.9827, 0.9755, 0.959}

\definecolor{kit-cyan}{RGB}{35, 161, 224}
\definecolor{kit-cyan100}{RGB}{35, 161, 224}
\definecolor{kit-cyan90}{rgb}{0.2235, 0.6682, 0.8906}
\definecolor{kit-cyan80}{rgb}{0.3098, 0.7051, 0.9027}
\definecolor{kit-cyan75}{rgb}{0.3529, 0.7235, 0.9088}
\definecolor{kit-cyan70}{rgb}{0.3961, 0.742, 0.9149}
\definecolor{kit-cyan60}{rgb}{0.4824, 0.7788, 0.9271}
\definecolor{kit-cyan50}{rgb}{0.5686, 0.8157, 0.9392}
\definecolor{kit-cyan40}{rgb}{0.6549, 0.8525, 0.9514}
\definecolor{kit-cyan30}{rgb}{0.7412, 0.8894, 0.9635}
\definecolor{kit-cyan25}{rgb}{0.7843, 0.9078, 0.9696}
\definecolor{kit-cyan20}{rgb}{0.8275, 0.9263, 0.9757}
\definecolor{kit-cyan15}{rgb}{0.8706, 0.9447, 0.9818}
\definecolor{kit-cyan10}{rgb}{0.9137, 0.9631, 0.9878}
\definecolor{kit-cyan5}{rgb}{0.9569, 0.9816, 0.9939}

\definecolor{kit-gray}{RGB}{0, 0, 0}
\definecolor{kit-gray100}{RGB}{0, 0, 0}
\definecolor{kit-gray90}{rgb}{0.1, 0.1, 0.1}
\definecolor{kit-gray80}{rgb}{0.2, 0.2, 0.2}
\definecolor{kit-gray75}{rgb}{0.25, 0.25, 0.25}
\definecolor{kit-gray70}{rgb}{0.3, 0.3, 0.3}
\definecolor{kit-gray60}{rgb}{0.4, 0.4, 0.4}
\definecolor{kit-gray50}{rgb}{0.5, 0.5, 0.5}
\definecolor{kit-gray40}{rgb}{0.6, 0.6, 0.6}
\definecolor{kit-gray30}{rgb}{0.7, 0.7, 0.7}
\definecolor{kit-gray25}{rgb}{0.75, 0.75, 0.75}
\definecolor{kit-gray20}{rgb}{0.8, 0.8, 0.8}
\definecolor{kit-gray15}{rgb}{0.85, 0.85, 0.85}
\definecolor{kit-gray10}{rgb}{0.9, 0.9, 0.9}
\definecolor{kit-gray5}{rgb}{0.95, 0.95, 0.95}

\ifCLASSOPTIONcompsoc
  \usepackage[caption=false,font=normalsize,labelfont=sf,textfont=sf]{subfig}
\else
  \usepackage[caption=false,font=footnotesize]{subfig}
\fi

\usepackage{cleveref}
\usepackage{csquotes} 

\usepackage{tikz}
\RequirePackage{tikz-dependency}
\usepackage{fontawesome5}
\usepackage{setspace}
\usetikzlibrary{calc}
\usetikzlibrary{arrows.meta, arrows}
\usetikzlibrary{positioning}
\usetikzlibrary{backgrounds}
\usetikzlibrary{fit}
\pgfdeclarelayer{foreground}
\pgfdeclarelayer{background}
\pgfsetlayers{background,depgroups,main,foreground}

\RequirePackage[proportional]{sourcesanspro}

\usepackage[theorems, skins, breakable, listings]{tcolorbox}
\usepackage{footnote}
\usepackage{etoolbox}
\usepackage{enumitem}
\tcbset{}

\newtcbtheorem[use counter=definition]{definition}{Definition}{%
	detach title,%
	colback=white,%
	colframe=kit-gray70,%
	fonttitle={\sffamily\bfseries\color{black}},%
    sharp corners=east,%
	leftrule=3mm,%
	subtitle style={%
		boxrule=0.4pt,%
		colback=kit-gray70,%
		fonttitle={\sffamily\bfseries\color{white}}%
	}%
}{def}

\crefname{definition}{Definition}{Definitionen}
\Crefname{definition}{Definition}{Definitionen}
\BeforeBeginEnvironment{definition}{\savenotes}
\AfterEndEnvironment{definition}{\spewnotes}

\newenvironment{highlightbox}[1]{
    \begin{tcolorbox}[title={#1},
    left=2mm,right=2mm,top=1mm,bottom=1mm]
    }{
    \end{tcolorbox}
}
\usepackage{eso-pic}
\begin{document}

\title{How Requirements Quality Makes (or Breaks) Traceability Link Recovery}

\author{
    \IEEEauthorblockN{Tobias Hey\IEEEauthorrefmark{1}}
    \IEEEauthorblockA{
        \textit{Karlsruhe Institute of Technology (KIT)}\\
        Karlsruhe, Germany \\
        hey@kit.edu}
    \and
    \IEEEauthorblockN{Julian Frattini\IEEEauthorrefmark{1}}
    \IEEEcompsocitemizethanks{\IEEEcompsocthanksitem\IEEEauthorrefmark{1}Both authors contributed equally to this study.}
    \IEEEauthorblockA{
        \textit{Chalmers University of Technology
and University of Gothenburg}\\
        Gothenburg, Sweden \\
        julian.frattini@chalmers.se}
}

\maketitle

\AddToShipoutPictureFG*{%
  \AtPageLowerLeft{%
    \raisebox{1.2cm}{%
      \hspace*{\dimexpr(\paperwidth-\textwidth)/2\relax}%
      \fbox{%
        \parbox{\dimexpr\textwidth-2\fboxsep-2\fboxrule\relax}{%
          \footnotesize
          ©2026 IEEE. Personal use of this material is permitted. Permission from IEEE must be obtained for all other uses, in any current or future media, including reprinting/republishing this material for advertising or promotional purposes, creating new collective works, for resale or redistribution to servers or lists, or reuse of any copyrighted component of this work in other works. DOI: (to be added once issued)
        }%
      }%
    }%
  }%
}

\begin{abstract}
    Traceability information between requirements and source code greatly benefits the maintenance of a software system.
    Since manually establishing \aclp{TL} is cumbersome and error-prone, previous research explored automated \ac{TLR} approaches to support this task.
    However, quality defects in requirements impact subsequent activities such as \ac{TLR}, yet evidence about this remains scarce.
    Our objective is to contribute empirical evidence on this impact.
    At the same time, we aim to understand how the performance of \ac{TLR} approaches varies given these quality defects.
    To this end, we annotated 28 types of quality defect in 189 use case descriptions from two datasets.
    Then, we executed five distinct \ac{TLR} approaches on the dataset and measured their performance in recovering \aclp{TL}.
    Finally, we performed statistical tests to quantify the defects' effect strength on this performance.
    Our results show that some quality defects harm \ac{TLR} performance, e.g., sentences that do not start with noun phrases, while others actually benefit performance, e.g., use cases that include implementation details.
    Moreover, different types of approaches respond differently to these defects.
    As a consequence, the performance-optimizing choice of a \ac{TLR} approach depends on the quality of the dataset.
\end{abstract}

\begin{IEEEkeywords}
traceability link recovery, requirements quality, quality defects, use case descriptions
\end{IEEEkeywords}

\section{Introduction}
\label{sec:intro}

\Acp{TL} connecting requirements to source code artifacts make explicit where a requirement is implemented.
These links greatly support the maintenance of a software system:
For example, if a requirement changes, the \acp{TL} locate which parts of the implementation need to be changed accordingly~\cite{cleland_software_2012}.
However, establishing \acp{TL} requires significant effort, is error-prone, and is rarely done upfront. 
Hence, automated \acl{TLR} approaches have become a popular avenue of research~\cite{charalampidou_empirical_2021}.

Recent approaches at automated \acf{TLR} have suggested that the quality of requirements artifacts may affect the performance of \ac{TLR}~\cite{hey2024requirements,vogelsang_impact_2025}.
For example, as illustrated in \autoref{fig:exampleUC}, sentences \sethlcolor{KITBlue!30!white}\hl{starting without a noun phrase} (i.e., lacking a semantic subject) may negatively influence the \ac{TLR} by not providing sufficient information on the actor of this activity.
The part that is on an \sethlcolor{KITOrange!30!white}\hl{inconsistent level of abstraction}, however, describes system-internal communication and not user-system interaction.
This additional implementation detail might positively influence the \ac{TLR} despite being commonly considered a defect~\cite{phalp2007assessing}, as the semantic gap to bridge between the artifacts is smaller.

\begin{figure}
    \centering
    \begin{tikzpicture}
        \node[draw, text width=0.97\columnwidth, align=left](req){
            \small
            \textbf{Use case name:} InsertFeedback\\
            \vspace{0.1cm}
            \textbf{Description:} Inserts a feedback for the selected site.\\
            \vspace{0.1cm}
            \textbf{Participating actor:} initialized by Tourist\\
            \vspace{0.1cm}
            \textbf{Entry conditions:} The tourist card is in a particular site.\\
            \vspace{0.1cm}
            \textbf{Flow of events User System:}
            \begin{enumerate}[label=\arabic*.]
                \item \sethlcolor{KITBlue!30!white}\hl{Activate} the feature on the issue of feedback. 
                \item \sethlcolor{KITBlue!30!white}\hl{Fill out} the form, selecting one vote and inserting a comment, then submit. 
                \item \sethlcolor{KITBlue!30!white}\hl{Confirm} the issue of feedback and \sethlcolor{KITBlue!30!white}\hl{insert} \sethlcolor{KITOrange!30!white}\hl{the selected site in the list of sites visited.} 
            \end{enumerate}
            \vspace{0.1cm}
            \textbf{Exit conditions:} The system shall notify the successful combi-\\
            \vspace{-0.1cm}
            nation of feedback to the site. 
            \vspace{0.1cm}
        };
    \end{tikzpicture}

    \caption{Use case description adapted from the eTour dataset with quality defects \sethlcolor{KITBlue!30!white}\hl{\emph{Starts without Noun Phrase}} and \sethlcolor{KITOrange!30!white}\hl{\emph{Inconsistent Level of Abstraction}}}
    \label{fig:exampleUC}
\end{figure}


While these assumptions seem plausible, we lack both a confirmation \emph{if} this effect is true and a quantification of \emph{how} strong it is.
This gap matters for several reasons.
From a \textbf{requirements quality perspective}, the lack of confirmation and quantification relegates the decision whether or not to address these alleged quality defects to mere guesswork instead of an evidence-based tradeoff.
If it were clear which quality defects have an actual, significant impact on \ac{TLR}, allocating resources towards removing them was justified.
From an \textbf{automated \ac{TLR} perspective}, the lack of confirmation and quantification obscures the eligibility and potential performance of \ac{TLR} algorithms.
If it were clear which quality defects have an actual, significant impact on \ac{TLR}, an organization could decide which \ac{TLR} algorithm fits the current status of their requirements best, what performance to expect, and which steps are necessary to improve it.

Reaching these two goals requires empirical evidence about the effect of requirements quality on \ac{TLR}.
We contribute the first step towards these goals via this empirical study answering the following research questions:

\begin{enumerate}[leftmargin=0cm,labelsep=0cm,align=left,label=\textbf{RQ\arabic*: },ref=RQ\arabic*]
    \item \label{RQ1}\emph{Which factors of requirements quality impact the performance of automated \acl{TLR}?}
    \item \label{RQ2}\emph{Do different approaches for automated \acl{TLR} respond differently to these factors?}
\end{enumerate}

To this end, we annotated 28 different types of requirements quality defects in 189 use case descriptions from two different datasets.
After executing five automated \ac{TLR} approaches to recover the \acp{TL} on all use case descriptions, we statistically determined whether any of these defects impact the performance of the \ac{TLR} approaches.
The resulting set of statistically significant defects indicates which defects are relevant to \ac{TLR} performance and whether an approach moderates their effect.

Thus, the contributions of this work are as follows:

\begin{enumerate}[leftmargin=0cm,labelsep=0cm,align=left,label=\textbf{C\arabic*: },ref=C\arabic*]
    \item \label{C1}\emph{A dataset of requirements quality defects in the use case descriptions of two projects.}
    \item \label{C2}\emph{An observational study on the impact of these quality defects on requirements to code \acl{TLR}.}
\end{enumerate}


\section{Related Work}
\label{sec:related}

Our work applies insights from requirements quality research to \acl{TLR} between requirements and code, briefly introduced in \Cref{sec:related:rq,sec:related:TLR}.

\subsection{Requirements Quality}
\label{sec:related:rq}

Requirements engineering is a means-to-an-end, i.e., it has the purpose of ``support[ing] the stakeholders in whatever activities they’re performing in the project''~\cite{femmer2018requirements}, including the activity of \ac{TLR}.
As means-to-an-end, the quality of a requirements artifact is determined by its impact on subsequent activities~\cite{femmer2015activities}.
This activity-based perspective on requirements quality properly frames requirements quality~\cite{frattini2023requirements}:
Only if a requirements quality defect has an impact on the attributes of a subsequent activity~\cite{frattini2024measuring} can it be considered an actual defect.
For example, if using \textit{passive voice} in a natural language requirements sentence produces a significant difference in the \textit{completeness} of the subsequent \textit{modeling} activity, then it can be considered a defect~\cite{femmer2014impact}.
On the other hand, if an alleged quality defect does not have a negative impact on any subsequent activity then this defect is negligible.

Recent roadmaps in requirements quality research have called for more empirical evidence on the relationship between requirements quality defects and potentially impacted activities~\cite{frattini2023requirements,femmer2018requirements}.
Since some alleged quality defects have been shown to have a positive impact on some activities~\cite{femmer2015activities}, recent publications suggested to refer to these with the more neutral term \textit{requirements quality factors} instead~\cite{frattini2022live}, which we adopt throughout this manuscript.

\subsection{Traceability Link Recovery}
\label{sec:related:TLR}

Automated \acl{TLR} between requirements and code has been studied since the 1990s, with early breakthroughs using \ac{IR} techniques. 
Methods such as \acp{VSM}~\cite{antoniol_recovering_2002}, \ac{LSI}~\cite{marcus_recovering_2003}, and \ac{LDA}~\cite{asuncion_software_2010} identify links based on textual similarity.

To improve performance, researchers have combined multiple \ac{IR} techniques~\cite{gethers_integrating_2011} and incorporated developer feedback with transitive linking, as seen in COMET~\cite{moran_improving_2020}. 
Another line of research investigated structural dependencies within source code, such as call and inheritance relationships, to further improve results~\cite{panichella_when_2013, kuang_can_2015}.

Gao et al. \cite{gao_using_2023,gao_triad_2024} utilize consensual biterms to refine \ac{TLR}, enriching artifacts with term pairs extracted from both requirements and code.
This approach improves performance with models like VSM and LSI.

Hey et al.~\cite{hey_improving_2021} demonstrate that a finer-grained analysis of artifacts can improve \ac{TLR}. 
Their approach, FTLR, establishes links at the sentence level for requirements and the method level for code, rather than using entire documents. 
FTLR captures the semantics of these smaller units through word embeddings. 
Additionally, they highlight the importance of selecting relevant information in the requirements to enhance performance and show that this selection can be automated using an \acs{LLM}-based classifier~\cite{hey2024requirements}.

\begin{figure*}
    \centering
    \includegraphics[clip,trim=1.3em 1.3em 1.3em 1.3em,width=0.9\linewidth]{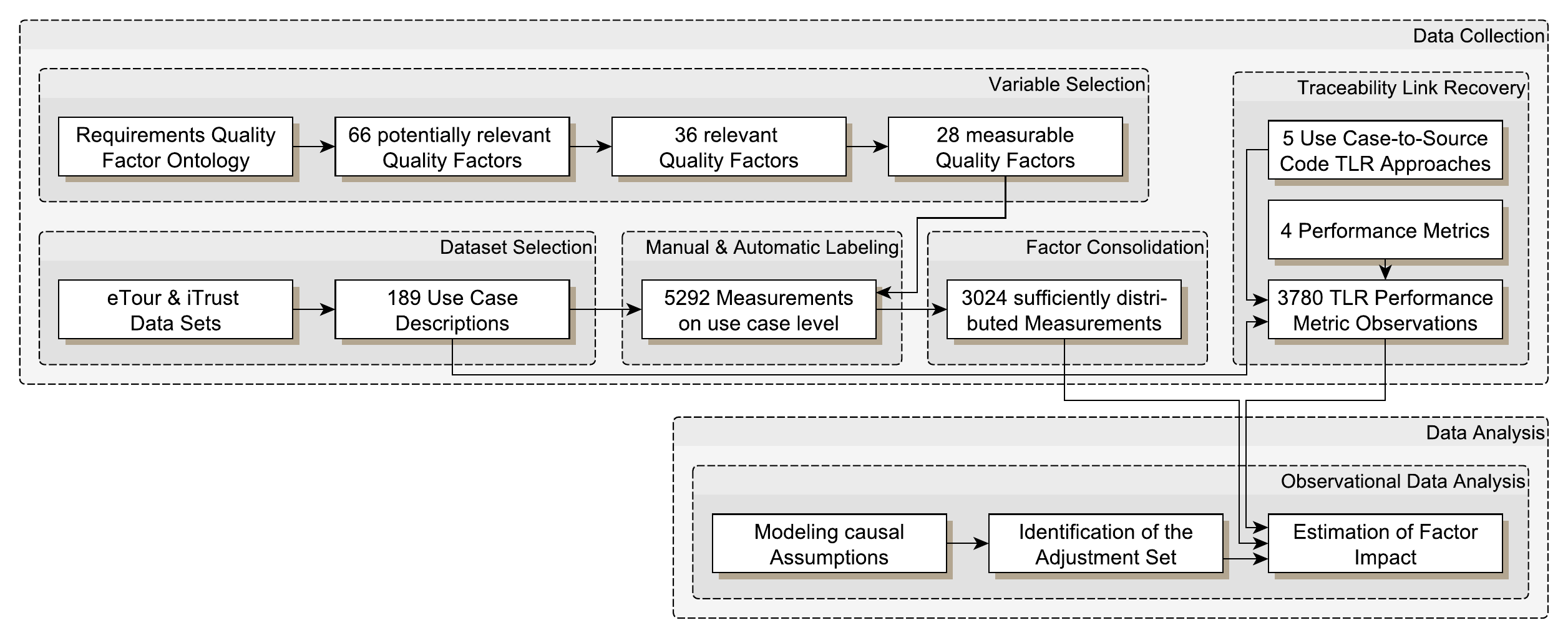}
    \caption{Overview of the phases of this study}
    \label{fig:study}
\end{figure*}

Despite these advances, traditional techniques struggle with semantically related but textually dissimilar artifacts. 
Recent approaches leverage \ac{ML} and \acp{LLM}, employing \acp{RNN}~\cite{guo_semantically_2017}, ranking models~\cite{wang_enhancing_2018}, active learning~\cite{mills_tracing_2019}, self-attention mechanisms~\cite{zhang_recovering_2021}, fine-tuned \acp{LLM}~\cite{lin_traceability_2021}, graph neural networks~\cite{wang_HGNNLink_2025}, and prompting techniques~\cite{rodriguez_prompts_2023,fuchss_lissa_2025,hey_requirements_2025, fuchss_beyond_2025} to bridge the semantic gap.
However, all but the prompt-based approaches require initial links of a project to be available to train the approach.

Fuchß et al.~\cite{fuchss_lissa_2025} show with their LiSSA framework that the combination of embedding-based retrieval as a candidate filter for a following LLM-based classification can be beneficial for requirements to code \ac{TLR}.

Most of these approaches do not consider the influence of requirements quality on \ac{TLR} or treat it as probabilistic noise~\cite{moran_improving_2020}.
To our knowledge, only the work by Vogelsang et al.~\cite{vogelsang_impact_2025} pursues a goal similar to ours.
They evaluate \acl{TLR} performance of two LLMs on five small, hand-crafted software systems, each providing both a clean and a defective version of the requirements specification, where the defective version was created by manually introducing five manually seeded defect types.
However, the authors note that these small systems (14-25 requirements and 141-220 LOC) may not be representative of real-world software, and that the deliberately selected defect types may not reflect those that occur in requirements specifications.
Therefore, they call for further empirical evidence. 
Our study responds to this call by providing complementary evidence under a different setup designed to address these limitations.

\section{Method}
\label{sec:method}

\ac{TLR} is valuable yet difficult to establish manually, and hence, benefits from automated support.
The quality of requirements artifacts likely influences the performance of these \ac{TLR} approaches~\cite{hey2024requirements}.
In this study, we aim to identify which factors of requirements quality affect \ac{TLR} and how strongly.
We identified occurrences of requirements quality factors in use case descriptions and observed their impact on the performance of several automated \ac{TLR} approaches.

\Cref{fig:study} visualizes the steps and phases in this study.
The group labels correspond to the respective sections and subsections in \Cref{sec:method}.

\subsection{Data Collection}
\label{sec:method:data}

Answering the RQs required a dataset with (1) requirements artifacts in a comparable, textual format, (2) source code artifacts, and (3) manually established \aclp{TL} between them. 
The latter served as the gold standard against which we evaluated the performance of a \ac{TLR} approach.

\subsubsection{Dataset Selection}
\label{sec:method:data:set}

\begin{table}
    \centering
    \caption{Overview of the used \acl{TLR} datasets illustrating the number of requirements (Req.), extracted sentences (Sent.), code files and trace links (Req. to Code)}
    \label{tab:dataset_overview}
    \footnotesize
    \begin{tabularx}{\columnwidth}{llrZZZZ}
        \toprule
                        &&& \multicolumn{4}{c}{Number of Artifacts}                    \\
        \cmidrule(lr){4-7}
        Dataset        &Domain& SLOC & Req. & Sent.& Code & TLs   \\
        \midrule
        eTour & Tourism & 12.4k                               & 58  & 266 &116 & 308  \\
        iTrust     &Healthcare & 14.6k                   & 131 & 333  &226 & 286  \\
        \midrule
        Total: & & & 189 & 599 & 342 & 594\\
        \bottomrule
    \end{tabularx}
\end{table}
\begin{table*}
    \centering
    \caption{List of selected quality factors with their origin, description, level, if they were measured manually (\faUsers) or automatically (\faCogs), are deterministically ({\fontfamily{qhv}\selectfont\textbf{D}}) or heuristically ({\fontfamily{qhv}\selectfont\textbf{H}}) measurable (\faMicroscope), and are sufficiently distributed (\faChartBar).}
    \label{tab:factors}
    \footnotesize
    \begin{tabularx}{\textwidth}{p{1.5em}p{3.8cm}cXccHHc}
        \toprule
         Lvl. & Name                                       &  Ref.               &  Description                                                                                                           & 
         \(\nicefrac{\text{\faUsers}}{\text{\faCogs}}\) & 
         \faMicroscope                              & 
         Heuristic                                  & 
         \faWrench                                  & 
         \faChartBar                                                                                                                                                                                                                                                                                                                                                                                                                                                                                                                                                                                                                       \\
        \midrule
        \multirow{20}{*}{%
        \cellcolor{white}
  \rotatebox[origin=c]{90}{use case level}%
}&Functional Duplication                                 & \cite{rago_approach_2014}      & If a use case includes the same functionality twice                                                                               & \faUsers & {\fontfamily{qhv}\selectfont\textbf{D}} &                                                                                                                                                                                                                                                & \textcolor{KITRed}{\faTimes}   & \textcolor{KITRed}{\faTimes} \\
        
        &\cellcolor{gray!20}Inputs or Outputs not Quantified                       & \cellcolor{gray!20}\cite{femmer2015activities}    & \cellcolor{gray!20}The degree to which quantitative inputs or outputs are specified in the use case                                                  & \cellcolor{gray!20}\faUsers & \cellcolor{gray!20}{\fontfamily{qhv}\selectfont\textbf{D}} &  \cellcolor{gray!20}                                                                                                                                                                                                                                              & \cellcolor{gray!20}\textcolor{KITRed}{\faTimes}   & \cellcolor{gray!20}\textcolor{KITRed}{\faTimes} \\
        &Use Case Naming Problems                               & \cite{ramos2009quality}        & Situations where the use case’s name bears no relation with the concept described or the same name is used for different concepts & \faUsers & {\fontfamily{qhv}\selectfont\textbf{H}}   & semantic similarity of a use case title to the use case content                                                                                                                                                                                & \textcolor{KITGreen}{\faCheck} & \textcolor{KITRed}{\faTimes} \\
       &\cellcolor{gray!20}Contains Justifications                                &\cellcolor{gray!20}\cite{parra_methodology_2015}  &\cellcolor{gray!20}Requirement contains justifications for decisions taken                                                                           &\cellcolor{gray!20}\faUsers &\cellcolor{gray!20}{\fontfamily{qhv}\selectfont\textbf{H}}   &\cellcolor{gray!20}                                                                                                                                                                                                                                               &\cellcolor{gray!20}\textcolor{KITRed}{\faTimes} &\cellcolor{gray!20}\textcolor{KITRed}{\faTimes} \\
        
        &Inappropriate Scope                                    & \cite{phalp2007assessing}      & A use case should only contain detail relevant to the problem statement                                                           & \faUsers & {\fontfamily{qhv}\selectfont\textbf{H}}   &                                                                                                                                                                                                                                                & \textcolor{KITRed}{\faTimes}   & \textcolor{KITRed}{\faTimes} \\
        & \cellcolor{gray!20}Incoherent Text Order                                  &\cellcolor{gray!20}\cite{phalp2007assessing}      &\cellcolor{gray!20}A use case should follow a logical path with events in the correct order                                                          &\cellcolor{gray!20}\faUsers &\cellcolor{gray!20}{\fontfamily{qhv}\selectfont\textbf{H}}   &\cellcolor{gray!20}                                                                                                                                                                                                                                               &\cellcolor{gray!20}\textcolor{KITRed}{\faTimes}   &\cellcolor{gray!20}\textcolor{KITRed}{\faTimes} \\
         &Mislocated Functional Requirement                                           & \cite{usdadiya_empirical_2019} & Functional requirements that are placed in fields addressing quality requirements                                                                           & \faUsers & {\fontfamily{qhv}\selectfont\textbf{H}}   &                                                                                                                                                                                                                                                & \textcolor{KITRed}{\faTimes}   & \textcolor{KITGreen}{\faCheck} \\
        & \cellcolor{gray!20}Free of Actor-Actor Interaction                        &\cellcolor{gray!20}\cite{usdadiya_empirical_2019} &\cellcolor{gray!20}Sentence should not contain actor to actor interaction                                                                            &\cellcolor{gray!20}\faUsers &\cellcolor{gray!20}{\fontfamily{qhv}\selectfont\textbf{H}}   &\cellcolor{gray!20}                                                                                                                                                                                                                                               &\cellcolor{gray!20}\textcolor{KITRed}{\faTimes}   &\cellcolor{gray!20}\textcolor{KITRed}{\faTimes} \\
        &Happy Use Case                                         & \cite{rago_approach_2014}      & If a use case does not contain a section for an alternative flow                                                                  & \faCogs  & {\fontfamily{qhv}\selectfont\textbf{D}} & existence of "alternative flow" in the use case                                                                                                                                                                                                & \textcolor{KITRed}{\faTimes}   & \textcolor{KITGreen}{\faCheck} \\
        & \cellcolor{gray!20}Large Use Case                                         &\cellcolor{gray!20}\cite{ramos2009quality}        &\cellcolor{gray!20}Many alternative flows and steps                                                                                                  &\cellcolor{gray!20}\faCogs  &\cellcolor{gray!20}{\fontfamily{qhv}\selectfont\textbf{D}} &\cellcolor{gray!20}number of steps and alternative flows                                                                                                                                                                                                          &\cellcolor{gray!20}\textcolor{KITRed}{\faTimes}   &\cellcolor{gray!20}\textcolor{KITGreen}{\faCheck} \\
        &Coherent                                               & \cite{phalp2007assessing}      & The sentence being written should repeat a noun in the last sentence or a previous sentence, if possible.                         & \faCogs  & {\fontfamily{qhv}\selectfont\textbf{D}} & Existence of a noun from the previous use case step in the next use case step                                                                                                                                                                  & \textcolor{KITRed}{\faTimes}   & \textcolor{KITGreen}{\faCheck} \\
         & \cellcolor{gray!20}Meaningless Actor                                      &\cellcolor{gray!20}\cite{rago_approach_2014}      &\cellcolor{gray!20}If an actor is wrongly defined or meaningless to the functionality                                                                &\cellcolor{gray!20}\faCogs  &\cellcolor{gray!20}{\fontfamily{qhv}\selectfont\textbf{H}}   &\cellcolor{gray!20}actor that appears in the actors field but not the main flow of the use case                                                                                                                                                                   &\cellcolor{gray!20}\textcolor{KITGreen}{\faCheck} &\cellcolor{gray!20}\textcolor{KITRed}{\faTimes} \\
        &Meaningless Use Case                                   & \cite{rago_approach_2014}      & Wrongly defined or meaningless use case that does not contribute to the requirements specification                                & \faCogs  & {\fontfamily{qhv}\selectfont\textbf{H}}   & use case without a gold-standard trace link to source code                                                                                                                                                                                     & \textcolor{KITGreen}{\faCheck} & \textcolor{KITRed}{\faTimes} \\
        & \cellcolor{gray!20}Tangled Requirements                                   &\cellcolor{gray!20}\cite{ramos2009quality}        &\cellcolor{gray!20}When a use case contains descriptions of several requirements or different functionalities                                        &\cellcolor{gray!20}\faCogs  &\cellcolor{gray!20}{\fontfamily{qhv}\selectfont\textbf{H}}   &\cellcolor{gray!20}Number of trace links from one requirement to multiple source code elements                                                                                                                                                                    &\cellcolor{gray!20}\textcolor{KITRed}{\faTimes}   &\cellcolor{gray!20}\textcolor{KITGreen}{\faCheck} \\
        &Scattered Requirements                                 & \cite{ramos2009quality}        & When the specification of one functionality is not encapsulated in one use case                                                   & \faCogs  & {\fontfamily{qhv}\selectfont\textbf{H}}   & Maximum number of incoming trace links of one of the source code files linked to the requirement                                                                                                                                               & \textcolor{KITRed}{\faTimes}   & \textcolor{KITGreen}{\faCheck} \\
        \midrule
        \multirow{10}{*}{\rotatebox[y=-8.5em]{90}{sentence level}}&Contains Alternatives                                  & \cite{phalp2007assessing}      & Alternative paths should be excluded from the main flow.                                                                          & \faUsers & {\fontfamily{qhv}\selectfont\textbf{D}} &                                                                                                                                                                                                                                                & \textcolor{KITRed}{\faTimes}   & \textcolor{KITGreen}{\faCheck} \\
        &\cellcolor{gray!20}Contains Clarifications                                &\cellcolor{gray!20}                               &\cellcolor{gray!20}Requirement contains clarifications of described functionalities                                                                  &\cellcolor{gray!20}\faUsers &\cellcolor{gray!20}{\fontfamily{qhv}\selectfont\textbf{H}}   &\cellcolor{gray!20}                                                                                                                                                                                                                                               &\cellcolor{gray!20}\textcolor{KITGreen}{\faCheck} &\cellcolor{gray!20}\textcolor{KITGreen}{\faCheck} \\
        &Inconsistent Level of Abstraction                      & \cite{phalp2007assessing}      & The use case should be at a consistent level of abstraction                                                                       & \faUsers & {\fontfamily{qhv}\selectfont\textbf{H}}   &                                                                                                                                                                                                                                                & \textcolor{KITRed}{\faTimes}   & \textcolor{KITGreen}{\faCheck} \\
        &\cellcolor{gray!20}Free of UI Design Details                              &\cellcolor{gray!20}\cite{femmer2015activities}    &\cellcolor{gray!20}The artifact should describe the problem domain instead of the solution domain                                                    &\cellcolor{gray!20}\faUsers &\cellcolor{gray!20}{\fontfamily{qhv}\selectfont\textbf{H}}   &\cellcolor{gray!20}                                                                                                                                                                                                                                               &\cellcolor{gray!20}\textcolor{KITRed}{\faTimes}   &\cellcolor{gray!20}\textcolor{KITGreen}{\faCheck} \\
       
        &Coordination Ambiguity                                 & \cite{ferrari_detecting_2018}  & The use of coordinating conjunctions (e.g., and or or) leads to multiple potential interpretations of a sentence                  & \faUsers & {\fontfamily{qhv}\selectfont\textbf{H}}   &                                                                                                                                                                                                                                                & \textcolor{KITRed}{\faTimes}   & \textcolor{KITRed}{\faTimes} \\
        
        &\cellcolor{gray!20}Anaphora                                               &\cellcolor{gray!20}\cite{ferrari_detecting_2018}  &\cellcolor{gray!20}A pronoun refers to a previous part of the text                                                                                   &\cellcolor{gray!20}\faCogs  &\cellcolor{gray!20}{\fontfamily{qhv}\selectfont\textbf{D}} &\cellcolor{gray!20}POS Tagging, only detect multiple potential noun phrases before pronoun                                                                                                                                                                        &\cellcolor{gray!20}\textcolor{KITRed}{\faTimes}   &\cellcolor{gray!20}\textcolor{KITGreen}{\faCheck} \\
        &Passive Voice                                          & \cite{ferrari_detecting_2018}  & Contains verb in passive voice                                                                                                    & \faCogs  & {\fontfamily{qhv}\selectfont\textbf{D}} & Dependency parser                                                                                                                                                                                                                              & \textcolor{KITRed}{\faTimes}   & \textcolor{KITGreen}{\faCheck} \\
        &\cellcolor{gray!20}Requirements Length                                    &\cellcolor{gray!20}\cite{ferrari_detecting_2018}  &\cellcolor{gray!20}Number of words in a use case step                                                                                      &\cellcolor{gray!20}\faCogs  &\cellcolor{gray!20}{\fontfamily{qhv}\selectfont\textbf{D}} &\cellcolor{gray!20}Token counting                                                                                                                                                                                                                                 &\cellcolor{gray!20}\textcolor{KITRed}{\faTimes}   &\cellcolor{gray!20}\textcolor{KITGreen}{\faCheck} \\
        &Starts without Noun Phrase                             & \cite{femmer_which_2017}       & Requirements shall start with the subject. As passive voice already covers this, we use noun phrases instead                      & \faCogs  & {\fontfamily{qhv}\selectfont\textbf{D}} & Syntax parsing                                                                                                                                                                                                                                 & \textcolor{KITRed}{\faTimes}   & \textcolor{KITGreen}{\faCheck} \\
       
        &\cellcolor{gray!20}Optional                                               &\cellcolor{gray!20}\cite{lami_automatic_2004}     &\cellcolor{gray!20}The sentence contains an optional part                                                                                            &\cellcolor{gray!20}\faCogs  &\cellcolor{gray!20}{\fontfamily{qhv}\selectfont\textbf{H}}   &\cellcolor{gray!20}Keyword search                                                                                                                                                                                                                                 &\cellcolor{gray!20}\textcolor{KITRed}{\faTimes}   &\cellcolor{gray!20}\textcolor{KITRed}{\faTimes} \\
        &Negatives                                              & \cite{femmer_rapid_2017}       & Statements of system capability not to be provided                                                                                & \faCogs  & {\fontfamily{qhv}\selectfont\textbf{H}}   & using negspaCy as a heuristic                                                                                                                                                                                                                  & \textcolor{KITRed}{\faTimes}   & \textcolor{KITGreen}{\faCheck} \\
        &\cellcolor{gray!20}Sentence Level Complexity                              &\cellcolor{gray!20}\cite{din_requirements_2008}   &\cellcolor{gray!20}A sentence level complexity metric called NPC-Sentence                                                                            &\cellcolor{gray!20}\faCogs  &\cellcolor{gray!20}{\fontfamily{qhv}\selectfont\textbf{H}}   &\cellcolor{gray!20}For each NP chunk, the occurrence count in a sentence is divided by the total occurrence counts in all sentences. Then all the frequency distributions of the NP chunks in the sentence are added together to form the final complexity value. &\cellcolor{gray!20}\textcolor{KITRed}{\faTimes}   &\cellcolor{gray!20}\textcolor{KITGreen}{\faCheck} \\
        &Complete Comparisons                              & \cite{hasso_detection_2019}    & A comparison is incomplete if there is no value for reference. E.g.: 'The system needs to be faster'.                             & \faCogs  & {\fontfamily{qhv}\selectfont\textbf{H}}   & Syntax parsing                                                                                                                                                                                                                                 & \textcolor{KITRed}{\faTimes}   & \textcolor{KITRed}{\faTimes} \\
        
        
        \bottomrule
    \end{tabularx}
\end{table*}

We aimed to gather a dataset that maximizes heterogeneity while also remaining reasonably comparable, i.e., in which requirements are specified in a comparable syntax.
To this end, we chose two datasets consisting of use case descriptions linked to source code commonly used in \ac{TLR} research: eTour and iTrust (see \autoref{tab:dataset_overview}).
Both were provided by the \emph{Center of Excellence for Software \& Systems Traceability} (CoEST) and map English use case descriptions to Java source code.
eTour includes use case descriptions following a template that demands the name, description, pre- and post-conditions, potential quality requirements, and the actual flow of events.
In eTour, entire use case descriptions are linked to the respective source code files that implement them.
In iTrust, the individual subflows of a use case description are mapped to the respective source code files.
As the dataset of iTrust did not include information on the use case name the subflow belongs to, we recovered that information from a version of the dataset used by Bencharrada et al.~\cite{bencharrada_supporting_2015}.
While these datasets may still not fully represent industry-grade software systems, they are at least several orders of magnitude more realistic in size than the systems previously studied~\cite{vogelsang_impact_2025}.

\subsubsection{Variable Selection}
\label{sec:method:data:variable}

Next, we determined the quality of each use case description.
We sampled requirements quality factors with an alleged impact on subsequent activities from the requirements quality factor ontology~\cite{frattini2022live}, an existing collection of such factors.
We selected all quality factors that apply to use cases, requirements, sentences, and phrases, which resulted in 66 potentially relevant factors.

Then, we jointly decided for each quality factor whether we hypothesize a potential (positive or negative) impact on the performance of \ac{TLR}, the response variable of interest in our study.
These decisions were based on the authors’ domain expertise in requirements engineering and traceability, as well as discussion among the authors until agreement was reached.
The following examples illustrate our decisions:

\begin{itemize}
    \item We \textbf{included} the quality factor \textit{Tangled Requirements}, i.e., that a use case should not describe several requirements or different functionalities~\cite{ramos2009quality}. We assume that tangled requirements make the recovery of all \acp{TL} more difficult and, thus, negatively impact the \ac{TLR} performance.
    \item We \textbf{included} the quality factor \textit{Free of UI Design Details}, i.e., that a requirement should not impose on the UI design~\cite{femmer2015activities}. While a violation of that factor is commonly considered a defect as it imposes on the solution space~\cite{frattini2025adopting}, we assume a positive impact of UI design details, since a \ac{TLR} approach might recognize these design details easier in the source code artifacts and, therefore, perform better.
    \item We \textbf{excluded} the quality factor \textit{Specification Clone} i.e., that there should not be a carbon copy of a requirement or a part thereof~\cite{juergens2010can}. While specification clones may have a negative impact on other activities (such as maintaining a set of requirements), we do not assume that it influences the performance of a \ac{TLR} approach.
\end{itemize}

Which factor to consider as relevant was a subjective decision.
To mitigate bias, we decided jointly after thorough discussion and documented choices for verification in our replication package~\cite{replication-package}.
After this step, we considered 36 of the 66 quality factors (54.5\%) relevant to \ac{TLR}.
We further analyzed these factors with regard to their operationalization, resulting in the final selection of factors displayed in \autoref{tab:factors}.

First, we determined the measurability of the eligible quality factors~\cite{femmer_which_2017}.
We distinguished between unmeasurable, manually measurable (\faUsers), and automatically measurable factors (\faCogs).
For example, we classified the quality factor \textit{viability of alternatives} as unmeasurable because it would require that ``[a]lternatives should be viable and make sense''~\cite{phalp2007assessing}.
Measuring this quality factor would require domain knowledge that was not accessible to us.
For each measurable quality factor, we further distinguished if we could measure it deterministically ((\faMicroscope~= {\fontfamily{qhv}\selectfont\textbf{D}})) or heuristically ((\faMicroscope~= {\fontfamily{qhv}\selectfont\textbf{H}})).

Excluding the 8 unmeasurable factors resulted in 28 (13 manually and 15 automatically) measurable requirements quality factors that we considered for the quality assessment of the use cases (see \autoref{tab:factors}).
Where quality factors were ambiguous, we specified the formulation to preserve as much information as possible while ensuring that the factor can be measured with reasonable precision.
For example, Ramos et al. consider a \textit{Large Use Case} when ``there are many alternative flows and steps''~\cite{ramos2009quality} without quantifying ``many''.
Instead, we consider the number of steps in a use case as its ``size'', thereby preserving more information.
Besides, we identified that most additional unnecessary information in the dataset were not ``justifications''~\cite{parra_methodology_2015} but ``clarifications'' such that we newly introduced the quality factor \textit{Contains Clarifications}.

As the quality factor \textit{Passive Voice} already covers passive phrase structures where the subject is not in the beginning of a sentence, we decided to change the quality factor \textit{Starting with Subject} to \textit{Starts without Noun Phrase}.
This allows us to measure the impact of non-standard phrase structures on \ac{TLR}.
Additionally, we changed the factor \textit{Free of NFRs} to \textit{Mislocated Functional Requirements}, as the only cases of non-functional requirements present in the datasets were quality requirements in the eTour dataset that were explicitly labeled in the use case template structure.
However, we identified cases where the statements in the quality requirement field were actually functional requirements, which led us to observe those cases of mislocated functional requirements instead.

Of the 28 factors, 10 were originally on sentence level and 18 on use case level according to the publications that proposed them~\cite{phalp2007assessing}.
We projected \textit{Contains Clarification}, \textit{Contains Alternatives}, and \textit{Inconsistent Level of Abstraction} to the sentence that causes the issue despite being originally defined as use case level factors 
which yields more precise insights.

\subsubsection{Manual Labeling}
\label{sec:method:data:manual}

To obtain data on the 13 manually measurable quality factors, the two authors manually labeled the two datasets.
As five of the 13 manually labeled factors were on sentence level, we labeled the 599 sentences of the datasets (see \autoref{tab:dataset_overview}) on this fine-grained level and the other eight factors on the 189 use case or subflow artifacts, resulting in 4507 data points. 
To ensure the reliability of this subjective step, we jointly developed an extraction guideline with instructions and examples, to be found in our replication package~\cite{replication-package}.
After agreeing on an initial version of the guidelines, each author individually labeled the dataset in three successive iterations. 
After each iteration, disagreements were discussed and the guidelines refined where necessary. 
Subsequently, each author individually revisited and relabeled all data points affected by the guideline changes to ensure consistency across the dataset.
Per iteration, we reached an average Cohen’s kappa of 85\%, 93\%, and 100\% for use case and 83\%, 100\% and 100\% for sentence level factors, confirming a substantial agreement~\cite{cohen1960coefficient}.

\subsubsection{Automatic Labeling}
\label{sec:method:data:automatic}

For the automatically measurable quality factors (\faCogs), we implemented analyzers using \ac{NLP} techniques.
The factors \emph{Happy Use Case}, \emph{Large Use Case}, \emph{Coherent}, \emph{Anaphora}, \emph{Passive Voice}, \emph{Starts without Noun Phrase}, and \emph{Requirements Length} we were able to measure deterministically (\faMicroscope~= {\fontfamily{qhv}\selectfont\textbf{D}}) or at least approximately whenever relying on NLP techniques.
For the other factors (\faMicroscope~= {\fontfamily{qhv}\selectfont\textbf{H}}) we had to resort to heuristics that can only provide indication for the existence of the factor.
We used the following heuristics: 

\emph{Meaningless Actor:} The actor appears in the actors field but not the main flow of the use case description.

\emph{Meaningless Use Case:} According to the traceability gold standard the use case has no links to the source code.

\emph{Tangled Requirements:} The number of trace links from one requirement to multiple source code elements.

\emph{Scattered Requirements:} The maximum number of incoming trace links of one of the code files linked to the requirement.
Note that the latter three heuristics are only operationalizations of their respective construct.
We assume meaningfulness, tangled-ness, and scattered-ness exist before any \acp{TL} and use the gold-standard \acp{TL} as proxies, since direct assessment would require domain knowledge beyond our reach.

The remaining heuristics are independent of the gold standard but rely on \ac{NLP} and keyword matching techniques, which might not cover all possible cases:

\emph{Optional:} Sentence contains one of the following keywords: \enquote{possibly}, \enquote{eventually}, \enquote{optionally}, \enquote{if possible}, \enquote{if appropriate}, \enquote{if needed}, \enquote{if necessary}, \enquote{if required}, \enquote{if applicable}, \enquote{if desired}, and \enquote{if applicable}

\emph{Negatives:} Sentence contains negating token (\enquote{not}, \enquote{n't}, \enquote{wouldn't}, \enquote{never}, \enquote{nowhere}, \enquote{noone}, \enquote{no-one}) and the negation has a dependency to a verb or auxiliary verb.

\emph{Sentence Level Complexity:} The sentence level complexity metric, or NPC-Sentence~\cite{din_requirements_2008}, can be calculated as follows. For each noun phrase (NP) chunk, the occurrence count in a sentence is divided by the total occurrence counts in all sentences. Then, all the frequency distributions of the NP chunks in the sentence are added together to form the final complexity value.

\emph{Complete Comparisons:} We consider a comparative adjective (JJR) or adverb (RBR) to be incomplete, if it is not followed by a comparison marker ("than" or "as") and is not part of a quantified conjunction (such as "four times faster").


\subsubsection{Factor Consolidation}
\label{sec:method:data:consolidation}

Two additional steps were necessary before utilizing the observations of quality defects in the use cases.
Since the gold standard \acp{TL} connect use cases with source code artifacts, we firstly needed to aggregate all sentence level factors per use case depending on their data type.
For Boolean sentence level variables (e.g., whether a sentence contains an \textit{Anaphora} or not) we calculated the ratio where the variable was \texttt{true}.
For numeric sentence level variables (e.g., the length of a step in a use case), we selected the maximum of all sentences per use case, as we deemed the maximum more expressive than the average.

Secondly, we excluded factors from the analysis if their natural distribution was strongly skewed.
For example, we did not encounter any instance of \textit{Functional Duplication} in our dataset.
Hence, all data points in our analysis had the same value for this variable (\textit{Functional Duplication} = 0) such that our observational data analysis would have not been able to make any inference about the influence of this factor.
We excluded 12 factors that were insufficiently distributed (\faChartBar~= \textcolor{KITRed}{\faTimes} in \autoref{tab:factors}), documented in our replication package~\cite{replication-package}.
While this meant discarding a considerable amount of manually gathered information, the distribution of the factors in our data was impossible to anticipate prior to their measurement.

\subsubsection{Traceability Link Recovery}
\label{sec:method:data:TLR}

We operationalized the \ac{TLR} task using several automatic algorithms tracing requirements to source code, since the different paradigms presented in \Cref{sec:related:TLR} may react differently to quality defects, thus acting as a confounding factor. 
We discard supervised \ac{ML}-based approaches, as they require a sufficient number of links per project for training, which may not be feasible given the size of our datasets. 
Furthermore, requiring existing links of a project limits the possible application scenarios for those approaches to scenarios where developers are available and willing to provide a substantial amount of those initial links~\cite{hey2024requirements}.
Requiring no initial trace links, so performing unsupervised \ac{TLR}, is applicable in all scenarios and, thus, broadens the applicability of the gained insights of our study.

The existing unsupervised \ac{TLR} approaches can be categorized by paradigms and used technologies (see \cref{sec:related:TLR}).
Most approaches are considering the \ac{TLR} as either an \acl{IR} or classification task.
Of the most common \ac{IR} paradigm-based approaches we chose the two most prevalent classical techniques \ac{VSM} and \ac{LSI} that use classical vector representations, and two approaches based on more modern language model-based embeddings FTLR, and LiSSA\_IR-only.
As representative of the more uncommon classification paradigm we make use of the recent LLM-based LiSSA approach.
This selection covers the relevant range of approaches presented in \Cref{sec:related:TLR}.

For \ac{VSM} and \ac{LSI} we make use of the implementation provided in the replication package of Gao et al.~\cite{gao_using_2023}.
FTLR~\cite{hey_improving_2021,hey2024requirements} presents the state-of-the-art of non-\ac{LLM}-based approaches.
It targets fine-grained (sentence level) information and combines \ac{IR}-based \ac{TLR} with neural network-based word embeddings.
We make use of FTLR's best performing configuration without use case template or requirements classification-based filters, as these filters are already targeted towards removing parts of the requirements with certain flaws.  
This configuration utilizes method comments and call dependencies.
For \ac{VSM}, \ac{LSI}, and FTLR we report the results with per-project optimized thresholds that were optimized based on F\textsubscript{1}-score on the respective gold standard.
Thus, the results represent the upper boundary of their performance.
LiSSA~\cite{fuchss_lissa_2025} is a \ac{RAG}–based framework that uses \acp{LLM} to recover trace links across different software artifacts such as requirements, documentation, and code. 
For requirements to code \ac{TLR} it outperforms FTLR, presenting the state-of-the-art in unsupervised \ac{TLR} for this artifact pairing.
Again, we use the best configuration reported by the authors, Fuchß et al., using GPT-4o with \ac{CoT} prompting.
Additionally, we can use only the retrieval part of LiSSA, which is basically an embedding-based \ac{IR} approach with OpenAI's \textit{text-embedding-3-large}, Fuchß et al. provided as well, which we call LiSSA\_IR-only.


As performance metrics we chose precision, recall, F\textsubscript{1}-score, and F\textsubscript{2}-score. 
Those metrics are commonly used in \ac{TLR} research~\cite{shin_guidelines_2015}.
They measure performance in a classification setting, providing insights on whether a link is correct or not. 
\Cref{tab:performance} lists the performance of all five investigated automated \ac{TLR} approaches in terms of the four metrics across the two sampled datasets.
To analyze the impact of the quality factors on performance, we computed precision, recall, F\textsubscript{1}-score, and F\textsubscript{2}-score at the use case level instead of the project level. 
Specifically, for each use case, we determined the associated true positives, false positives, true negatives, and false negatives and derived the corresponding metrics from these counts.

\begin{table}
    \caption{\textbf{P}recision, \textbf{R}ecall, and F-scores of used \ac{TLR} approaches}
    \centering
    \setlength\tabcolsep{4.2pt}
    \footnotesize
    \begin{tabularx}{\columnwidth}{lZZZZZZZZ}
        \toprule
        & \multicolumn{4}{c}{eTour} & \multicolumn{4}{c}{iTrust} \\
        \cmidrule(lr){2-5}\cmidrule(lr){6-9}
        Approach& P. & R. & F\textsubscript{1} & F\textsubscript{2} & P. & R. & F\textsubscript{1} & F\textsubscript{2} \\
        \midrule
        VSM & .557 & .427 & .483 & .448 & .208 & .227 & .217 & .223 \\
        LSI & .452 & .453 & .453 & .453 & .251 & .255 & .253 & .254 \\
        FTLR & .505 & .597 & .548 & .576 & .234 & .241 & .238 & .240 \\
        LiSSA\_IR-only & .216 & .815 & .342 & .525 & .058 & .531 & .105 & .202 \\
        LiSSA & .409 & .734 & .526 & .633 & .199 & .451 & .276 & .360 \\

        \bottomrule
    \end{tabularx}

    \label{tab:performance}
\end{table}

\subsection{Data Analysis}
\label{sec:method:analysis}

To answer our RQs, we performed regression analyses where we regress our outcome variables on the requirements quality factors.
We followed the Pearlian framework for statistical causal inference~\cite{pearl2009causality,pearl2009causal} as summarized by Siebert~\cite{siebert2023applications} using Bayesian data analysis (BDA) methods~\cite{mcelreath2018statistical}.
While BDA is less established in SE research, it produces more granular inferences from the data, preserves uncertainty instead of reducing complex data to binary results, and makes decisions on causal assumptions and statistical modeling explicit~\cite{furia2019bayesian,mcelreath2018statistical}.
As we cannot provide a comprehensive, pedagogical introduction to BDA methods in this manuscript, we refer the interested reader to appropriate textbooks~\cite{mcelreath2018statistical}, examples~\cite{furia2019bayesian,furia2022applying,torkar2020bayesian,frattini2025applying}, and our replication package~\cite{replication-package}.


The framework for statistical causal inference dictates three steps: modeling, identification, and estimation.
We apply the same framework for answering both research questions.

\paragraph{Modeling}
First, we created a causal model in the form of a directed, acyclic graph (DAG)~\cite{elwert2013graphical}.
In this causal DAG, every node represents a variable and every edge an assumed causal relationship.
More importantly, the absence of any edge between two variables represents a certain lack of a relationship between those variables.
Our causal DAG consists of all (measurable) variables with an edge directed to the outcome \textit{\ac{TLR} performance}.
There is only one interrelation among these variables: if a sentence is written in \textit{Passive Voice}, we also assume that it \textit{Starts without Noun Phrase}, which we encode with a directed edge.
Additionally, we assume that the \textit{dataset} itself has a direct influence both on \ac{TLR} performance, because one dataset might just be easier to trace than the other, and on all quality factors, because the distribution of quality factors varies between the datasets.
Similarly, we assume that the \textit{TLR approach} influences \ac{TLR} performance, as different paradigms for automated \acl{TLR} may handle quality defects differently well.

\paragraph{Identification}
Next we derive a statistical model from the aforementioned causal model~\cite{pearl2009causality}.
In this identification step, we selected all variables relevant to our investigation, as well as all variables necessary to de-confound the effect of interest~\cite{cinelli2024crash}.
In our case, this requires the inclusion of the \textit{dataset} variable in our statistical model, as it is a potential common cause of all independent variables (i.e., different datasets may have different distributions of the variables) and the dependent variables (i.e., different datasets may naturally perform better or worse in \ac{TLR}).

\paragraph{Estimation}
Finally, we estimated the strength of effect that each independent variable has on the response---i.e., dependent---variables through a regression model.
Within the Bayesian data analysis methods, we first selected an appropriate distribution type for each response variable~\cite{gelman2020bayesian}.
We selected the distribution types based on the maximum entropy criterion~\cite{jaynes2003probability} and ontological assumptions.
As the values of all metrics are bounded within $[0, 1]$, we chose a distribution from the \texttt{Beta} family~\cite{paolino2001maximum}.
The Beta distribution family deals with the heteroskedasticity of the data at the edges of the $[0, 1]$ range.
To account for occurrences of the values 0 and 1 in each metric, we selected the \textit{zero-one-inflated Beta distribution}~\cite{ospina2012general}.

Next, we defined regression formulae for each response variable.
These specify each response variable in relationship to all available independent variables.
The strength and direction of the effect of each variable---the quantity of interest to answer our research question---is represented by a coefficient in the form of a random variable.

We selected uninformative prior distributions for the coefficients of each included predictor and confirmed their eligibility via prior predictive checks~\cite{wesner2021choosing}.
After confirming their eligibility, we trained each model with the collected data.
Hamiltonian Monte Carlo Markov Chains (MCMC)~\cite{brooks2011handbook} update the coefficient distributions based on the empirical data.
During this process, the parameters of the coefficient distributions are adjusted to better predict the response variable based on the independent variables~\cite{mcelreath2018statistical}.
After the training process, we perform posterior predictive checks, which work similarly to the prior predictive check but use the updated posterior coefficient distributions instead of the prior distributions. 

To evaluate each model, we inspected the distributions of the updated path coefficients.
Traditionally, coefficients where the 95\% credibility interval (CI) is consistent with 0, i.e., where the CI overlaps with 0, are not considered statistically significant~\cite{frattini2025applying}.
To avoid the fallacy of reducing complex data down to the binary property of statistical significance, we still investigate all factors whose coefficient shows a notable effect.
Therefore, in addition to the statistically significant factors 
we consider coefficients where the 50\% credibility interval is not consistent with 0 to show a weak impact.

While the posterior path coefficients answer RQ1, their interpretation is challenging since the regression formulae are wrapped by a \texttt{logit}-link function~\cite{mcelreath2018statistical}.
This link function scales the summed value of all regressors to the range $(0, 1)$, which is the appropriate range for the estimated main parameter in a Beta distribution.
To aid understanding the actual impact of significant quality factors on the response variables we additionally plot selected \textit{marginal effects}.
Marginal effects visualize the change in the response variable when manipulating one isolated regressor while keeping all others at a representative level~\cite{mize2019general}.
Numeric predictors were fixed at their mean, and categorical predictors at their mode.
In our case, the marginal effects were generated for FTLR assuming the dataset eTour.
As such, they show the actual effect of this variable on the scale of the response variable.

To answer RQ2, we performed the same data analysis process but with additional regression formulae.
We formulated one regression model for every quality factor, containing an \textit{interaction effect} between this factor and the \ac{TLR} approach.
Such an interaction effect discerns whether a quality factor affects the response variable differently for different \ac{TLR} approaches~\cite{denters1989conditional}.
We only used the F\textsubscript{2}-score as a response instead of all four performance metrics to limit the space of potential models.
We plotted the conditional effects to visualize similar and different effects of quality factors across \ac{TLR} approaches, which provides insight into whether some approaches deal with certain significant factors better or worse than others.

\section{Results}
\label{sec:results}

\begin{figure*}
    \centering
    \includegraphics[trim=0 17pt 5pt 4pt, clip, width=0.89\linewidth]{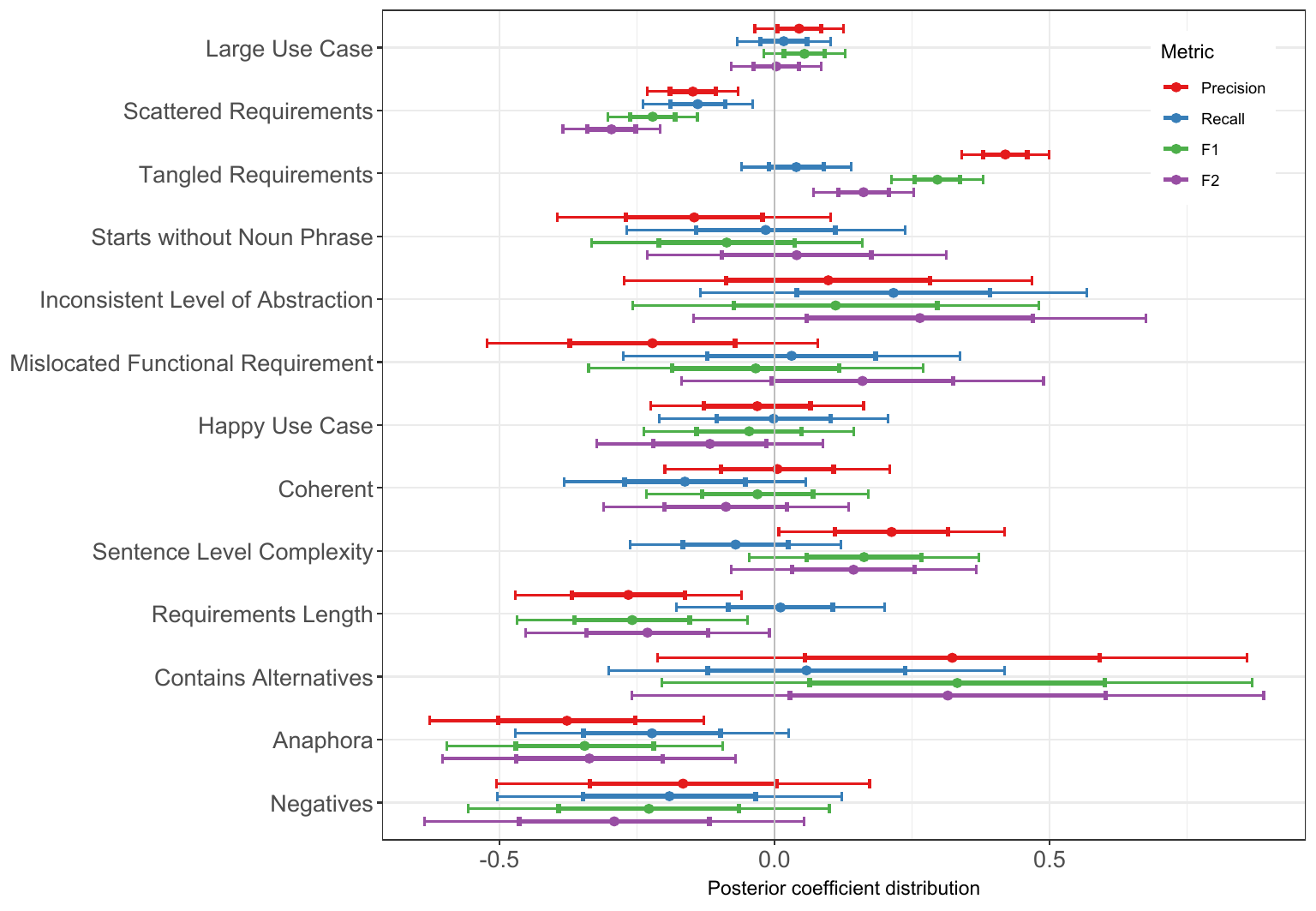}
    \caption{Posterior coefficient distribution of (at least weakly) significant coefficients from the analysis}
    \label{fig:obs:coef}
\end{figure*}

In this section, we show the results from the data analysis and set them into context of existing requirements quality work. 
For conciseness, we report only relevant and meaningful results.
Our replication package contains the data, material, and figures of all omitted results~\cite{replication-package}.

\subsection{Posterior Coefficient Distributions}
\label{sec:results:posteriors}

\Cref{fig:obs:coef} visualizes the posterior distributions of variable coefficients from the data analysis that were at least weakly significant.
The dots represent the average effect of the factor on the respective response variable (coded in colors), while the two whiskers represent the single and double standard deviation distance from the average.
The further the dot is away from 0, the stronger the effect.
The more narrow the whiskers, the more certain the effect.

Of the 16 investigated quality factors, 13 show at least a weakly significant effect on at least one of the four outcome metrics.
Some factors exhibit narrow CIs, suggesting a very certain estimated effect.
\textit{Large Use Cases} affect only precision and the F\textsubscript{1}-score slightly, where \textit{Scattered Requirements} impede all metrics significantly.
\textit{Tangled Requirements}, on the other hand, have a positive effect on all metrics but recall.

The other factors have less certain effects on the response variables.
Notable are the negative effects of use cases with a high \textit{Requirements Length}, and the ratio of sentences that contain \textit{Anaphora} and \textit{Negatives} on the performance of \ac{TLR} approaches.
On the other hand, the degree to which a requirement is written on an \textit{Inconsistent Level of Abstraction} and \textit{Contains Alternatives} will benefit particularly recall and F\textsubscript{2}-score.
Sentence-level complexity significantly benefits precision at a minor cost to recall, improving F\textsubscript{1}- and F\textsubscript{2}-scores overall.
The maximum \textit{Requirements Length} (i.e., the highest number of words used in a single use case step), on the other hand, harms precision significantly, as does the existence of anaphora and negatives, which additionally also affect recall.

Putting these results into context, some quality factors behave as expected. 
For example, the fact that a sentence that \textit{Starts without a Noun Phrase} omits information potentially relevant to \ac{TLR} and, hence, impedes \ac{TLR} performance aligns with our expectations.
Similarly, the negative effect of \textit{Requirements Length} on all metrics but recall aligns with literature suggesting that larger use cases are harder to parse~\cite{ramos2009quality}.
Recommendations against \textit{Anaphora}~\cite{ferrari_detecting_2018} and \textit{Negatives}~\cite{femmer_rapid_2017} align with the observed negative effect on automated \ac{TLR}.

However, some quality factors traditionally labeled as \enquote{defects} exhibit a contrary effect.
This is most notable for \textit{Inconsistent Level of Abstraction}, \textit{Contains Alternatives}, and \textit{Tangled Requirements}.
Requirements specified on an inconsistent level of abstraction, i.e., containing system-internal information, are traditionally considered solution-orientated~\cite{fernandez2012field}.
Such requirements that impose on the solution-space rather than fully specifying the problems a system is supposed to solve have been shown to negatively affect the downstream software development process.
They cause requirements engineers to commit to premature design decisions at an early stage of the project that require costly rework later~\cite{frattini2025adopting}.
Hence, literature advocates against an \textit{Inconsistent Level of Abstraction} and considers it a quality defect~\cite{phalp2007assessing}.
Yet, for automated \ac{TLR}, this factor shows a slight positive effect, possibly because internal implementation cues aid in recovering links to code artifacts.

A similar case holds for \textit{Contains Alternatives}: while
established quality guidelines recommend separating alternative steps from the main scenario of a use case for clarity~\cite{phalp2007assessing}, automated \ac{TLR} appears to benefit from blended scenario information.
Again, labeling this quality factor as a \enquote{defect} may not be warranted in the case of automated \ac{TLR}.

The positive effect of \textit{Sentence Level Complexity} on precision and, consequently, F\textsubscript{1}- and F\textsubscript{2}-scores is less intuitive to interpret.
Prior literature assumes complex sentences have a negative effect, particularly in hindering human comprehension~\cite{din_requirements_2008}.
However, our results suggest this does not translate to automated \ac{TLR} approaches.
Due to the operationalization of this factor using a suggested compound metric~\cite{din_requirements_2008}, though, we cannot trace this effect to specific linguistic properties.
Contrasting the effect of the individual components of the complexity metric might provide further insights in the future.

\begin{highlightbox}{Answer to RQ1}
    13 out of 16 quality factors show at least a weakly significant effect on at least one outcome metric.
    Factors such as scattered requirements, requirements length, and the use of negatives impede \ac{TLR} performance, while factors such as tangled requirements or contains alternatives benefit it.
    Overall, these results show that requirements quality affects automated \ac{TLR} approaches, although the strength and direction of the effects vary across factors.
\end{highlightbox}

\subsection{Marginal Effects}
\label{sec:results:marginal}
\begin{figure*}
    \centering
    \includegraphics[trim=5pt 5pt 5pt 5pt, clip,width=\linewidth]{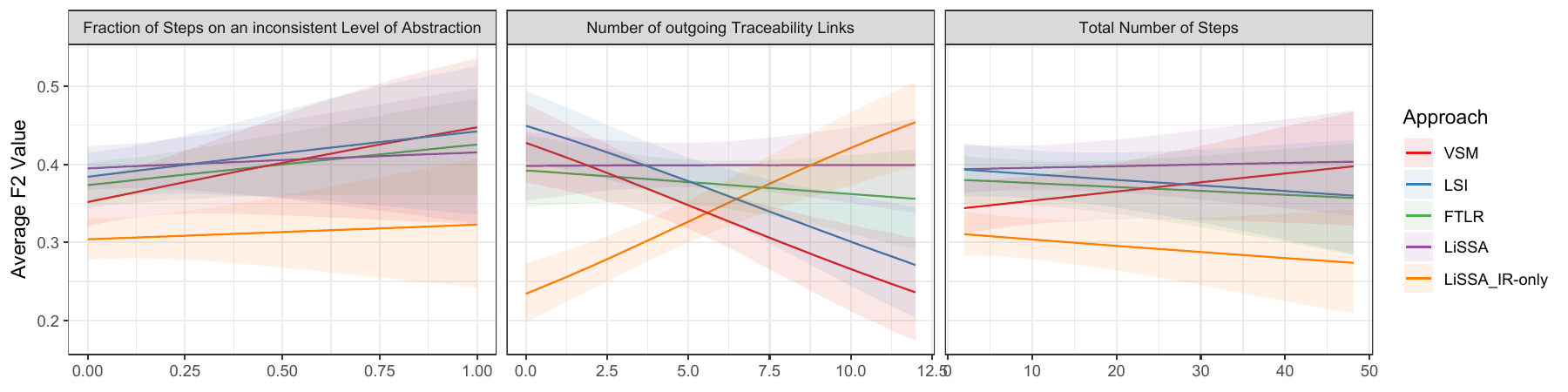}
    \caption{Conditional effects between the approach and significant quality factors}
    \label{fig:conditional}
\end{figure*}
\begin{figure}
    \centering
    \includegraphics[trim=5pt 12pt 2pt 5pt, clip, width=\linewidth]{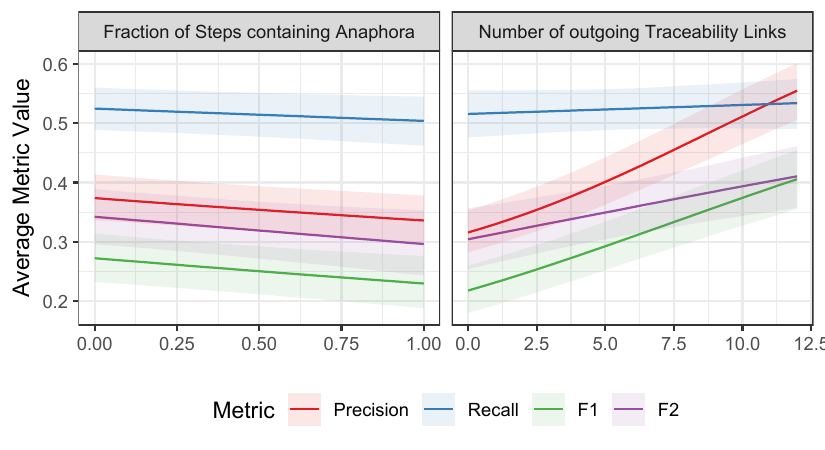}
    \caption{Marginal effects on the outcome metrics}
    \label{fig:marginal}
\end{figure}

\Cref{fig:marginal} visualizes the marginal effects of two selected factors, \textit{Anaphora} and \textit{Tangled Requirements}, on the outcome metrics.
Due to space limitations, we sampled these two as they show different factor scales (percentage vs. absolute), different effect directions (negative vs. positive effect), and also different effect strengths (moderate vs. strong).
These two act as representative visualizations of the translation from posterior coefficient distributions to actual metric outcomes.
Other marginal effects can be found in our replication package~\cite{replication-package}.

While the x-axes of both plots show the different levels of the two factors, the two y-axes show the expected average value (as a line) and the 95\% CI around it.
Both plots show that recall, while highest, is affected the least and, therefore, shows the flattest slope, which is consistent with the coefficient values in \Cref{fig:obs:coef} where these intersect zero.
On the other hand, the significant values of the posterior coefficients for \textit{Tangled Requirements} in terms of precision, F\textsubscript{1}- and F\textsubscript{2}-score translate to steep slopes across the range of possible values of the quality factor.
On average, increasing the number of outgoing links from a (theoretical) minimum to the maximum entails an increase in precision by about 0.2.

\begin{highlightbox}{Answer to RQ1}
    The strength of effects that quality factors exhibit on the performance on automated \ac{TLR} approaches varies.
    At the extreme, use cases with a high number of outgoing traceability links (12, as observed in our sample) achieve about 20\% higher precision than use cases with the minimum number of outgoing traceability links.
\end{highlightbox}

\subsection{Conditional Effects}
\label{sec:results:conditional}


\Cref{fig:conditional} visualizes selected conditional effects.
The x-axes represent the different levels of the quality factors (one percentaged, one real), the y-axis the average expected F\textsubscript{2}-score.
For each \ac{TLR} approach, we plot a colored line (mean) and a 50\% CI as a transparent ribbon around it (the width of the CI is arbitrarily chosen and intended only to convey estimation uncertainty without cluttering the figure).
For conciseness, we sampled representative effects for this manuscript.
The conditional effects of all omitted significant quality factors are available in our replication package~\cite{replication-package}.

Most regression models suggest that \ac{TLR} approaches respond similarly to varying levels of a given quality factor.
For example, the left subplot visualizes the conditional effect of an increasing degree of \textit{Inconsistent Level of Abstraction} on the F\textsubscript{2}-score per \ac{TLR} approach.
As seen in \Cref{fig:obs:coef}, this quality factor has a slight, yet not significant, positive impact on the F\textsubscript{2}-score.
All five approaches respond similarly: as use case steps increasingly include design- or implementation-specific details, F\textsubscript{2}-scores tend to improve.
The different intercepts of the five lines represent the different baseline performances of the five approaches, in line with \Cref{tab:performance} which identifies LiSSA\_IR-only as the automated \ac{TLR} approach that performs lowest in terms of F\textsubscript{2}-score.
The slightly different slopes represent the interaction effects of the approaches to the quality factor.
Notably, VSM benefits most (i.e., has the steepest slope), likely because it relies entirely on keyword matching, and implementation-specific terminology aids its \ac{TLR}.
Thus, while \textit{Inconsistent Level of Abstraction} may rightly be labeled a quality defect for downstream design activities~\cite{phalp2007assessing,frattini2025adopting}, it does not hinder but rather aids automated \ac{TLR}.

Only a few quality factors show markedly divergent effects across approaches.
The middle subplot of \Cref{fig:conditional} visualizes the conditional effect between \textit{Tangled Requirements} and the used approach on the F\textsubscript{2}-score.
While FTLR and LiSSA remain fairly stable, VSM and LSI performance suffers the more outgoing \aclp{TL} connect a requirement with source code artifacts.
On the other hand, LiSSA\_IR-only strongly benefits from that same increase in the quality factor.
The interpretation of this effect is less straight-forward.
Requirements quality literature suggests that a ``use case that contains several different functionalities could be hard to understand''~\cite{ramos2009quality}, which intuitively translates also to a negative impact on automated \ac{TLR} performance.
However, \Cref{fig:obs:coef} and \Cref{fig:marginal} show a significant positive effect on precision, which also affects the F\textsubscript{1}- and F\textsubscript{2}-score.
The conditional effect offers a finer-grained explanation.
LiSSA\_IR-only responds positively as this approach retrieves the top 20 target artifacts, where 20 is, by design, a generous upper bound considering the two selected datasets. 
Thus, the requirements that do contain many outgoing \aclp{TL} to source code artifacts improve the precision of LiSSA\_IR-only as a larger portion of the retrieved 20 artifacts are likely to be true positives.
LiSSA is stable because it additionally classifies the top 20 potential target artifacts retrieved from the IR approach~\cite{fuchss_lissa_2025}, and thus, enhances its precision independently from the number of outgoing links.
VSM and LSI, are threshold-optimized for F\textsubscript{1}-score, and as a lower number of outgoing links is more common in the dataset they respond negatively to the quality factor exceeding the derived threshold.
FTLR is also threshold optimized~\cite{hey2024requirements} but constructs links from the code files, making it less sensitive to this factor.
Thus, whether \textit{Tangled Requirements} constitute a defect for automated \ac{TLR} depends on the approach: effects vary drastically across methods.

The right subplot containing the conditional effect between \textit{Large Use Cases} and the approaches shows a less pronounced, yet opposite, response.
While LiSSA remains stable, the F\textsubscript{2}-score of FTLR, LSI, and LiSSA\_IR-only decreases the more steps a use case contains, while the score of VSM increases.
This divergence can be interpreted as follows.
LiSSA, powered by GPT-4o, demonstrates the strongest capability to interpret semantic content across multiple steps, enabling it to retrieve relevant links regardless of the number of steps in a use case.
VSM’s performance improves as use cases grow longer, likely because its TF-IDF-based matching benefits from increased term frequency and lexical coverage:
longer text provides more opportunities for keyword overlap with source code artifacts, even when those artifacts relate to semantically distinct steps.
In contrast, the other IR-based methods such as LSI and LiSSA\_IR-only exhibit slight performance degradation with increasing step count, as their vector-space embeddings become less precise when aggregating semantically heterogeneous content across multiple steps.

\begin{highlightbox}{Answer to RQ2}
    For most studied quality factors, all five automated \ac{TLR} approaches respond similarly. 
    However, the approach notably moderates the effect of some quality factors on the F\textsubscript{2}-score.
    This shows automated \ac{TLR} approaches respond differently to certain quality factors such as the number of outgoing trace links and total number of steps contained in the use case depending on their design and architecture.
\end{highlightbox}

\section{Threats to Validity}
\label{sec:discussion:threats}

Our study is subject to the following threats to validity, categorized according to the types by Wohlin et al.~\cite{wohlin2012experimentation}.

\subsection{Threats to internal validity}
We limited our selection of independent variables to those contained in an existing repository of requirements quality factors~\cite{frattini2022live}.
While this makes for a transparent selection based on established previous research, it also risks ignoring other potentially relevant variables not included in the repository.
Secondly, we had to exclude several variables as described in \Cref{sec:method:data:variable} because we lacked the domain knowledge to collect them.
The only mitigation strategy within our means was to make our causal assumptions explicit in a causal DAG documented in our replication package~\cite{replication-package}.
This DAG makes explicit which variables we considered and which we ignored or had to discard.
This allows other scholars to scrutinize our decisions and propose a competing causal DAG with revised assumptions as typical in principled statistical causal inference via model comparison~\cite{mcelreath2018statistical}.

\subsection{Threats to external validity}
The generalizability of the inferred results is limited by the inclusion of only two datasets and five \ac{TLR} approaches in our analysis.
While further datasets and approaches would increase the external validity of the results, our strict selection criteria mentioned in \Cref{sec:method:data:set,sec:method:data:TLR} limited the population to sample from.
We used all data and working approaches available to us in the study but acknowledge that an extrapolation of the results, especially from only two datasets, has to be considered with caution.

\subsection{Threats to conclusion validity}
We obtained our conclusions via Bayesian methods in a statistical causal inference framework.
Every design decision, such as the selection of the distribution type to model the response variables, represents a potential threat to the validity of our conclusion.
To minimize this threat, we jointly discussed all these decisions and made them to the best of our knowledge under the guidance of established literature~\cite{elwert2013graphical,jaynes2003probability,mcelreath2018statistical} and documented every subjective influence in our replication package for other scholars to assess.
Still, improvements to the data analysis via even more complex regression models (e.g., involving hierarchical models) may further improve the conclusion validity.
Finally, the interaction effects used to answer RQ2 are known to demand a much larger volume of data in order to converge on a result~\cite{gelman2018interaction}.
To acknowledge this threat, we plot the CIs in the visualization of conditional effects as seen in \Cref{fig:conditional} to retain explicit information about the uncertainty of these effects.

\subsection{Threats to construct validity}
Several quality factors from the repository~\cite{frattini2022live} were described only vaguely and were difficult to measure.
Our operationalizations may affect how well we could represent intended concepts.
For example, the heuristics for \textit{Scattered Requirements}, \textit{Tangled Requirements}, and \textit{Meaningless Use Case} are based on the \ac{TL} gold standard of the datasets.
Thus, the indicators depend on the quality of the gold standard, as well.
For example, if the gold standard is incomplete, more use cases may be considered \emph{Meaningless Use Cases} than actually are.
Additionally, the heuristics for \emph{Tangled} and \emph{Scattered Requirements} must be considered with caution, as a high number of trace links does not necessarily mean that these defects actually are present.
For example, if many classes are needed to implement one use case step, that does not necessarily mean that this step has tangled requirements.
We operationalized vague quality factors to the best of our knowledge after extensive discussions and documented all decisions to allow iterative improvement of the results' construct validity.

\section{Discussion}
\label{sec:discussion}

Finally, we discuss the obtained results and put them into context with regard to their implications for \ac{TLR} (\Cref{sec:discussion:implicationsTLR}) and requirements quality (\Cref{sec:discussion:implicatiosnReqQual}), and outline potential ways of continuing this research in \Cref{sec:discussion:future}.

\subsection{Implications for TLR}
\label{sec:discussion:implicationsTLR}

Besides contributing evidence to previously postulated hypotheses, our results enable the three use cases mentioned in \Cref{sec:intro}.
Firstly, they allow predicting the fitness of a requirements specification for the activity of automated \ac{TLR}. 
An organization planning to recover \aclp{TL} between use cases and source code artifacts using one of the studied approaches can evaluate their use cases with the list of quality factors identified as significant.
This quality assessment allows estimating how well the use cases in their current form will allow automated \ac{TLR}.
Secondly, the results can inform the design of requirements writing guidelines that instruct on producing \enquote{high}-quality requirements artifacts.
Instead of basing writing guidelines on intuition, anecdotal evidence, or theory alone, our results serve as support for the particular effect of quality factors on automated \ac{TLR}.
An organization aiming to maximize the performance of an automated \ac{TLR} approach can advise writing use cases in particular ways, e.g., avoiding excessive \textit{Requirements Length} (for certain \ac{TLR} approaches) or the use of \textit{Negatives}.
Such writing guidelines are more likely to produce requirements that can be considered of high-quality for the particular task than those not based on such evidence~\cite{femmer2018requirements,frattini2023requirements}.
Finally, the results can inform the design of automatic \ac{TLR} algorithms.
As acknowledged by the authors, the performance of even state-of-the-art approaches (\Cref{tab:performance}) leaves room for improvement which harms their practical applicability~\cite{hey2024requirements,fuchss_lissa_2025}.
Considering the impact of requirements quality in the design of new automated \ac{TLR} approaches may allow further improving their performance.

\subsection{Implications for Requirements Quality}
\label{sec:discussion:implicatiosnReqQual}

Our results support the activity-based paradigm that the quality of requirements artifacts impacts the performance of subsequent activities~\cite{femmer2015activities,frattini2023requirements}, i.e., the automated \ac{TLR}.
As such, this study provides empirical evidence requested by research roadmaps for requirements quality~\cite{femmer2018requirements,frattini2023requirements}, and therefore, contributes to the larger endeavor of evidence-based quality assessment of requirements artifacts.
Accumulating sufficient evidence will allow companies to reason about which alleged quality factor to address (e.g., resolving all \textit{anaphora}) and which to ignore (e.g., keeping \textit{alternatives} in use case descriptions) for a given activity like automatic \ac{TLR}.

When using requirements artifacts for multiple activities---as usually the case---the quantification of the impact of a quality factor on each activity allows deciding the trade-off.
For example, when deciding whether or not to consider \textit{Inconsistent Levels of Abstraction} as a defect worth removing, comparing the benefit that they have on automated \ac{TLR} with the drawback they have on design specifications~\cite{frattini2025adopting} allows for an evidence-based decision.


\subsection{Future Work}
\label{sec:discussion:future}

The results invite several streams of future work.
For one, extrapolating the base assumption that the quality of artifacts impacts the activities in which they are used, an obvious extension of our work would be to also consider how the quality of the source code artifacts to which the requirements are traced impacts the \ac{TLR} approach.
Future studies could extend our approach to find (1) effects of the source code quality and (2) interaction effects between the quality factors of the two artifact types.
Furthermore, there are several other types of requirements artifacts as well as traceability tasks involving requirements artifacts, such as inter-requirements \ac{TLR} or requirements to model \ac{TLR}.
Since use case descriptions are mostly textual, we anticipate that our results may look similar for other textual requirements artifacts.
Extending the approach to other \ac{TLR} tasks will yield a more complete picture of the impact of requirements quality, though.

\section{Conclusion}
\label{sec:conclusion}

The quality of requirements artifacts affects the performance of automated \ac{TLR} approaches trying to connect them to source code that implements these requirements. 
In this study, we were able to show that use case descriptions with high degrees of sentences not beginning with noun phrases, long requirements, or scattered requirements will impede the performance of some automated \ac{TLR} approaches.
On the other hand, inconsistent levels of abstraction and tangled requirements might benefit them.

We recommend that the results of our study will be used as starting points for context-specific investigations on the impact of any of these factors.
Ultimately, once the validity of our conclusions is strengthened, we envision that this avenue of research can inform organizations about how fit for automated \acl{TLR} their requirements artifacts are.

\section*{Data Availability Statement}

All protocols, data, source code, analysis scripts, figures, and results are available at \url{https://github.com/JulianFrattini/rq4tlr} and archived in our replication package~\cite{replication-package}.

\section*{Acknowledgements}
This work was funded by Core Informatics at KIT (KiKIT) of the Helmholtz Assoc. (HGF) and supported by the German Research Foundation (DFG) - SFB 1608 - 501798263 and KASTEL Security Research Labs.
Additionally, this work was supported by the KKS foundation through the S.E.R.T. Research Profile project at Blekinge Institute of Technology.
The authors express their deep gratitude to Michael Unterkalmsteiner for providing feedback on the manuscript.

\IEEEtriggeratref{53}
\bibliographystyle{IEEEtran}
\bibliography{material/references}

@inproceedings{hey2024requirements,
  title     = {Requirements Classification for Traceability Link Recovery},
  author    = {Hey, Tobias and Keim, Jan and Corallo, Sophie},
  booktitle = {2024 IEEE 32nd International Requirements Engineering Conference (RE'24)},
  year      = {2024},
  doi       = {10.1109/RE59067.2024.00024}
}

@inproceedings{hey_requirements_2025,
author="Hey, Tobias
and Fuch{\ss}, Dominik
and Keim, Jan
and Koziolek, Anne",
editor="Hess, Anne
and Susi, Angelo",
title={{Requirements Traceability Link Recovery via Retrieval-Augmented Generation}},
booktitle="Requirements Engineering: Foundation for Software Quality",
year="2025",
publisher="Springer Nature Switzerland",
address="Cham",
pages="381--397",
isbn="978-3-031-88531-0"
}

@article{frattini2023requirements,
  title     = {Requirements quality research: a harmonized theory, evaluation, and roadmap},
  author    = {Frattini, Julian and Montgomery, Lloyd and Fischbach, Jannik and Mendez, Daniel and Fucci, Davide and Unterkalmsteiner, Michael},
  journal   = {Requirements Engineering},
  pages     = {1--14},
  year      = {2023},
  publisher = {Springer},
  doi       = {10.1007/s00766-023-00405-y}
}

@inproceedings{frattini2022live,
  title        = {A live extensible ontology of quality factors for textual requirements},
  author       = {Frattini, Julian and Montgomery, Lloyd and Fischbach, Jannik and Unterkalmsteiner, Michael and Mendez, Daniel and Fucci, Davide},
  booktitle    = {2022 IEEE 30th International Requirements Engineering Conference (RE)},
  pages        = {274--280},
  year         = {2022},
  organization = {IEEE},
  doi          = {10.1109/RE54965.2022.00041}
}

@inproceedings{frattini2024measuring,
  title        = {Measuring the Fitness-for-Purpose of Requirements: An initial Model of Activities and Attributes},
  author       = {Frattini, Julian and Fischbach, Jannik and Fucci, Davide and Unterkalmsteiner, Michael and Mendez, Daniel},
  booktitle    = {2024 IEEE 30th International Requirements Engineering Conference (RE)},
  year         = {2024},
  organization = {IEEE},
  doi          = {10.1109/RE59067.2024.00047}
}

@article{femmer2018requirements,
  title     = {Requirements quality is quality in use},
  author    = {Femmer, Henning and Vogelsang, Andreas},
  journal   = {IEEE Software},
  volume    = {36},
  number    = {3},
  pages     = {83--91},
  year      = {2018},
  publisher = {IEEE},
  doi       = {10.1109/MS.2018.110161823}
}

@inproceedings{femmer2015activities,
  title        = {It's the activities, stupid! a new perspective on RE quality},
  author       = {Femmer, Henning and Mund, Jakob and Fern{\'a}ndez, Daniel M{\'e}ndez},
  booktitle    = {2015 IEEE/ACM 2nd International Workshop on Requirements Engineering and Testing},
  pages        = {13--19},
  year         = {2015},
  organization = {IEEE},
  doi          = {10.1109/RET.2015.11}
}

@inproceedings{juergens2010can,
  title     = {Can clone detection support quality assessments of requirements specifications?},
  author    = {Juergens, Elmar and Deissenboeck, Florian and Feilkas, Martin and Hummel, Benjamin and Schaetz, Bernhard and Wagner, Stefan and Domann, Christoph and Streit, Jonathan},
  booktitle = {Proceedings of the 32nd ACM/IEEE International Conference on Software Engineering-Volume 2},
  pages     = {79--88},
  year      = {2010}
}

@inproceedings{ramos2009quality,
  title        = {Quality improvement for use case model},
  author       = {Ramos, Ricardo and Castro, Jaelson and Alencar, Fernanda and Ara{\'u}jo, Jo{\~a}o and Moreira, Ana and da Computacao, C de Engenharia and Penteado, R},
  booktitle    = {2009 XXIII Brazilian Symposium on Software Engineering},
  pages        = {187--195},
  year         = {2009},
  organization = {IEEE}
}

@article{phalp2007assessing,
  title     = {Assessing the quality of use case descriptions},
  author    = {Phalp, Keith Thomas and Vincent, Jonathan and Cox, Karl},
  journal   = {Software Quality Journal},
  volume    = {15},
  pages     = {69--97},
  year      = {2007},
  publisher = {Springer}
}

@article{antoniol_recovering_2002,
  title    = {Recovering Traceability Links between Code and Documentation},
  author   = {Antoniol, G. and Canfora, G. and Casazza, G. and Lucia, A. De and Merlo, E.},
  year     = {2002},
  month    = oct,
  journal  = {IEEE Transactions on Software Engineering},
  volume   = {28},
  number   = {10},
  pages    = {970--983},
  issn     = {0098-5589},
  doi      = {10.1109/TSE.2002.1041053}
}

@inproceedings{asuncion_software_2010,
  title     = {Software Traceability with Topic Modeling},
  booktitle = {2010 {{ACM}}/{{IEEE}} 32nd {{International Conference}} on {{Software Engineering}}},
  author    = {Asuncion, H. U. and Asuncion, A. U. and Taylor, R. N.},
  year      = {2010},
  month     = may,
  volume    = {1},
  pages     = {95--104},
  doi       = {10.1145/1806799.1806817}
}

@inproceedings{fuchss_lissa_2025,
  author={Fuchß, Dominik and Hey, Tobias and Keim, Jan and Liu, Haoyu and Ewald, Niklas and Thirolf, Tobias and Koziolek, Anne},
  booktitle={2025 IEEE/ACM 47th International Conference on Software Engineering (ICSE)}, 
  title={{LiSSA}: {Toward Generic Traceability Link Recovery Through Retrieval-Augmented Generation}}, 
  year={2025},
  volume={},
  number={},
  pages={1396-1408},
  doi={10.1109/ICSE55347.2025.00186}
}

@INPROCEEDINGS{fuchss_beyond_2025,
  author={Fuchß, Dominik and Schwedt, Stefan and Keim, Jan and Hey, Tobias},
  booktitle={2025 IEEE 33rd International Requirements Engineering Conference Workshops (REW)}, 
  title={{Beyond Retrieval: A Study of Using LLM Ensembles for Candidate Filtering in Requirements Traceability}}, 
  year={2025},
  volume={},
  number={},
  pages={5-12},
  doi={10.1109/REW66121.2025.00006}
}

@inproceedings{gao_triad_2024,
  title      = {{{TRIAD}}: {{Automated Traceability Recovery}} Based on {{Biterm-enhanced Deduction}} of {{Transitive Links}} among {{Artifacts}}},
  shorttitle = {{{TRIAD}}},
  booktitle  = {Proceedings of the {{IEEE}}/{{ACM}} 46th {{International Conference}} on {{Software Engineering}}},
  author     = {Gao, Hui and Kuang, Hongyu and Assun{\c c}{\~a}o, Wesley K. G. and {Mayr-Dorn}, Christoph and Rong, Guoping and Zhang, He and Ma, Xiaoxing and Egyed, Alexander},
  year       = {2024},
  month      = apr,
  series     = {{{ICSE}} '24},
  pages      = {1--13},
  publisher  = {Association for Computing Machinery},
  address    = {New York, NY, USA},
  doi        = {10.1145/3597503.3639164},
  urldate    = {2025-02-19},
  isbn       = {9798400702174}
}

@inproceedings{gao_using_2023,
  title     = {Using {{Consensual Biterms}} from {{Text Structures}} of {{Requirements}} and {{Code}} to {{Improve IR-Based Traceability Recovery}}},
  booktitle = {Proceedings of the 37th {{IEEE}}/{{ACM International Conference}} on {{Automated Software Engineering}}},
  author    = {Gao, Hui and Kuang, Hongyu and Sun, Kexin and Ma, Xiaoxing and Egyed, Alexander and M{\"a}der, Patrick and Rong, Guoping and Shao, Dong and Zhang, He},
  year      = {2023},
  month     = jan,
  series    = {{{ASE}} '22},
  publisher = {Association for Computing Machinery},
  address   = {New York, NY, USA},
  doi       = {10.1145/3551349.3556948},
  isbn      = {978-1-4503-9475-8}
}

@inproceedings{gethers_integrating_2011,
  title     = {On Integrating Orthogonal Information Retrieval Methods to Improve Traceability Recovery},
  booktitle = {2011 27th {{IEEE International Conference}} on {{Software Maintenance}} ({{ICSM}})},
  author    = {Gethers, M. and Oliveto, R. and Poshyvanyk, D. and Lucia, A. D.},
  year      = {2011},
  month     = sep,
  pages     = {133--142},
  doi       = {10.1109/ICSM.2011.6080780}
}

@inproceedings{guo_semantically_2017,
  title     = {Semantically {{Enhanced Software Traceability Using Deep Learning Techniques}}},
  booktitle = {Proceedings of the 39th {{International Conference}} on {{Software Engineering}}},
  author    = {Guo, Jin and Cheng, Jinghui and {Cleland-Huang}, Jane},
  year      = {2017},
  series    = {{{ICSE}} '17},
  pages     = {3--14},
  publisher = {IEEE Press},
  address   = {Piscataway, NJ, USA},
  doi       = {10.1109/ICSE.2017.9},
  urldate   = {2018-06-06},
  isbn      = {978-1-5386-3868-2}
}

@inproceedings{hey_improving_2021,
  title     = {Improving {{Traceability Link Recovery Using Fine-grained Requirements-to-Code Relations}}},
  booktitle = {2021 {{IEEE International Conference}} on {{Software Maintenance}} and {{Evolution}} ({{ICSME}})},
  author    = {Hey, Tobias and Chen, Fei and Weigelt, Sebastian and Tichy, Walter F.},
  year      = {2021},
  month     = sep,
  pages     = {12--22},
  issn      = {2576-3148},
  doi       = {10.1109/ICSME52107.2021.00008}
}

@article{kuang_can_2015,
  title    = {Can {{Method Data Dependencies Support}} the {{Assessment}} of {{Traceability Between Requirements}} and {{Source Code}}?},
  author   = {Kuang, Hongyu and M{\"a}der, Patrick and Hu, Hao and Ghabi, Achraf and Huang, LiGuo and L{\"u}, Jian and Egyed, Alexander},
  year     = {2015},
  month    = nov,
  journal  = {J. Softw. Evol. Process},
  volume   = {27},
  number   = {11},
  pages    = {838--866},
  issn     = {2047-7473},
  doi      = {10.1002/smr.1736},
  urldate  = {2018-10-24}
}

@inproceedings{lin_traceability_2021,
  title = {Traceability {{Transformed}}: {{Generating}} More {{Accurate Links}} with {{Pre-Trained BERT Models}}},
  shorttitle = {Traceability {{Transformed}}},
  booktitle = {Proceedings of the 43rd {{International Conference}} on {{Software Engineering}}},
  author = {Lin, Jinfeng and Liu, Yalin and Zeng, Qingkai and Jiang, Meng and {Cleland-Huang}, Jane},
  year = {2021},
  month = nov,
  series = {{{ICSE}} '21},
  pages = {324--335},
  publisher = {IEEE Press},
  address = {Madrid, Spain},
  issn = {1558-1225},
  doi = {10.1109/ICSE43902.2021.00040},
  urldate = {2025-03-11},
  isbn = {978-1-4503-9085-9}
}

@inproceedings{marcus_recovering_2003,
  title     = {Recovering {{Documentation-to-source-code Traceability Links Using Latent Semantic Indexing}}},
  booktitle = {Proceedings of the 25th {{International Conference}} on {{Software Engineering}}},
  author    = {Marcus, Andrian and Maletic, Jonathan I.},
  year      = {2003},
  series    = {{{ICSE}} '03},
  pages     = {125--135},
  publisher = {IEEE Computer Society},
  address   = {Washington, DC, USA},
  urldate   = {2018-10-23},
  isbn      = {978-0-7695-1877-0}
}

@inproceedings{mills_tracing_2019,
  title      = {Tracing with {{Less Data}}: {{Active Learning}} for {{Classification-Based Traceability Link Recovery}}},
  shorttitle = {Tracing with {{Less Data}}},
  booktitle  = {2019 {{IEEE International Conference}} on {{Software Maintenance}} and {{Evolution}} ({{ICSME}})},
  author     = {Mills, Chris and {Escobar-Avila}, Javier and Bhattacharya, Aditya and Kondyukov, Grigoriy and Chakraborty, Shayok and Haiduc, Sonia},
  year       = {2019},
  month      = sep,
  pages      = {103--113},
  issn       = {1063-6773},
  doi        = {10.1109/ICSME.2019.00020}
}

@inproceedings{moran_improving_2020,
  title     = {Improving the Effectiveness of Traceability Link Recovery Using Hierarchical Bayesian Networks},
  booktitle = {Proceedings of the {{ACM}}/{{IEEE}} 42nd {{International Conference}} on {{Software Engineering}}},
  author    = {Moran, Kevin and Palacio, David N. and {Bernal-C{\'a}rdenas}, Carlos and McCrystal, Daniel and Poshyvanyk, Denys and Shenefiel, Chris and Johnson, Jeff},
  year      = {2020},
  month     = jun,
  series    = {{{ICSE}} '20},
  pages     = {873--885},
  publisher = {Association for Computing Machinery},
  address   = {New York, NY, USA},
  doi       = {10.1145/3377811.3380418},
  urldate   = {2021-01-05},
  isbn      = {978-1-4503-7121-6}
}

@inproceedings{panichella_when_2013,
  title     = {When and {{How Using Structural Information}} to {{Improve IR-Based Traceability Recovery}}},
  booktitle = {2013 17th {{European Conference}} on {{Software Maintenance}} and {{Reengineering}}},
  author    = {Panichella, A. and McMillan, C. and Moritz, E. and Palmieri, D. and Oliveto, R. and Poshyvanyk, D. and Lucia, A. De},
  year      = {2013},
  month     = mar,
  pages     = {199--208},
  doi       = {10.1109/CSMR.2013.29}
}

@inproceedings{rodriguez_prompts_2023,
  title      = {Prompts {{Matter}}: {{Insights}} and {{Strategies}} for {{Prompt Engineering}} in {{Automated Software Traceability}}},
  shorttitle = {Prompts {{Matter}}},
  booktitle  = {2023 {{IEEE}} 31st {{International Requirements Engineering Conference Workshops}} ({{REW}})},
  author     = {Rodriguez, Alberto D. and Dearstyne, Katherine R. and {Cleland-Huang}, Jane},
  year       = {2023},
  month      = sep,
  pages      = {455--464},
  issn       = {2770-6834},
  doi        = {10.1109/REW57809.2023.00087},
  urldate    = {2024-09-24}
}

@inproceedings{wang_enhancing_2018,
  title     = {Enhancing {{Automated Requirements Traceability}} by {{Resolving Polysemy}}},
  booktitle = {2018 {{IEEE}} 26th {{International Requirements Engineering Conference}} ({{RE}})},
  author    = {Wang, W. and Niu, N. and Liu, H. and Niu, Z.},
  year      = {2018},
  month     = aug,
  pages     = {40--51},
  doi       = {10.1109/RE.2018.00-53}
}

@article{wang_HGNNLink_2025,
  title = {{{HGNNLink}}: Recovering Requirements-Code Traceability Links with Text and Dependency-Aware Heterogeneous Graph Neural Networks},
  shorttitle = {{{HGNNLink}}},
  author = {Wang, Bangchao and Zou, Zhiyuan and Liang, Xuanxuan and Jin, Huan and Liang, Peng},
  year = {2025},
  month = may,
  journal = {Autom Softw Eng},
  volume = {32},
  number = {2},
  pages = {55},
  issn = {1573-7535},
  doi = {10.1007/s10515-025-00528-2},
  urldate = {2025-08-12},
  langid = {english},
 }

@inproceedings{zhang_recovering_2021,
  title     = {Recovering {{Semantic Traceability}} between {{Requirements}} and {{Source Code Using Feature Representation Techniques}}},
  booktitle = {2021 {{IEEE}} 21st {{International Conference}} on {{Software Quality}}, {{Reliability}} and {{Security}} ({{QRS}})},
  author    = {Zhang, Meng and Tao, Chuanqi and Guo, Hongjing and Huang, Zhiqiu},
  year      = {2021},
  month     = dec,
  pages     = {873--882},
  issn      = {2693-9177},
  doi       = {10.1109/QRS54544.2021.00096}
}

@article{bencharrada_supporting_2015,
  title    = {Supporting Requirements Update during Software Evolution},
  author   = {Ben Charrada, Eya and Koziolek, Anne and Glinz, Martin},
  year     = {2015},
  month    = mar,
  journal  = {J. Softw. Evol. Process},
  volume   = {27},
  number   = {3},
  pages    = {166--194},
  issn     = {2047-7473},
  doi      = {10.1002/smr.1705},
  urldate  = {2025-02-19}
}

@inproceedings{shin_guidelines_2015,
  title     = {Guidelines for {{Benchmarking Automated Software Traceability Techniques}}},
  booktitle = {2015 {{IEEE}}/{{ACM}} 8th {{International Symposium}} on {{Software}} and {{Systems Traceability}}},
  author    = {Shin, Yonghee and Hayes, Jane Huffman and {Cleland-Huang}, Jane},
  year      = {2015},
  month     = may,
  pages     = {61--67},
  issn      = {2157-2194},
  doi       = {10.1109/SST.2015.13}
}

@book{din_requirements_2008,
  title     = {Requirements content goodness and complexity measurement based on NP chunks},
  author    = {Din, Chao Y and Rine, D},
  year      = {2008},
  publisher = {VDM Publishing Saarbr{\"u}cken}
}

@book{pearl2009causality,
  title     = {Causality},
  author    = {Pearl, Judea},
  year      = {2009},
  publisher = {Cambridge university press}
}

@article{pearl2009causal,
  title   = {Causal inference in statistics: An overview},
  author  = {Pearl, Judea},
  year    = {2009},
  journal = {Statistical Surveys},
  doi     = {10.1214/09-SS057}
}

@article{siebert2023applications,
  title     = {Applications of statistical causal inference in software engineering},
  author    = {Siebert, Julien},
  journal   = {Information and Software Technology},
  volume    = {159},
  pages     = {107198},
  year      = {2023},
  publisher = {Elsevier}
}

@book{mcelreath2018statistical,
  title     = {Statistical rethinking: A Bayesian course with examples in R and Stan},
  author    = {McElreath, Richard},
  year      = {2018},
  publisher = {Chapman and Hall/CRC}
}

@article{furia2019bayesian,
  title     = {Bayesian data analysis in empirical software engineering research},
  author    = {Furia, Carlo A and Feldt, Robert and Torkar, Richard},
  journal   = {IEEE Transactions on Software Engineering},
  volume    = {47},
  number    = {9},
  pages     = {1786--1810},
  year      = {2019},
  publisher = {IEEE}
}

@article{furia2022applying,
  title     = {Applying Bayesian analysis guidelines to empirical software engineering data: The case of programming languages and code quality},
  author    = {Furia, Carlo A and Torkar, Richard and Feldt, Robert},
  journal   = {ACM Transactions on Software Engineering and Methodology (TOSEM)},
  volume    = {31},
  number    = {3},
  pages     = {1--38},
  year      = {2022},
  publisher = {ACM New York, NY}
}

@article{torkar2020bayesian,
  title     = {Bayesian data analysis in empirical software engineering: The case of missing data},
  author    = {Torkar, Richard and Feldt, Robert and Furia, Carlo A},
  journal   = {Contemporary Empirical Methods in Software Engineering},
  pages     = {289--324},
  year      = {2020},
  publisher = {Springer}
}

@article{cinelli2024crash,
  title     = {A crash course in good and bad controls},
  author    = {Cinelli, Carlos and Forney, Andrew and Pearl, Judea},
  journal   = {Sociological Methods \& Research},
  volume    = {53},
  number    = {3},
  pages     = {1071--1104},
  year      = {2024},
  publisher = {Sage Publications Sage CA: Los Angeles, CA},
  doi       = {10.1177/00491241221099552}
}

@article{gelman2020bayesian,
  title   = {Bayesian workflow},
  author  = {Gelman, Andrew and Vehtari, Aki and Simpson, Daniel and Margossian, Charles C and Carpenter, Bob and Yao, Yuling and Kennedy, Lauren and Gabry, Jonah and B{\"u}rkner, Paul-Christian and Modr{\'a}k, Martin},
  journal = {arXiv preprint arXiv:2011.01808},
  year    = {2020}
}

@book{jaynes2003probability,
  address   = {Cambridge},
  author    = {Jaynes, E. T.},
  publisher = {Cambridge University Press},
  title     = {Probability theory: {T}he logic of science},
  year      = 2003
}

@article{wesner2021choosing,
  title     = {Choosing priors in {B}ayesian ecological models by simulating from the prior predictive distribution},
  author    = {Wesner, Jeff S and Pomeranz, Justin PF},
  journal   = {Ecosphere},
  volume    = {12},
  number    = {9},
  pages     = {e03739},
  year      = {2021},
  publisher = {Wiley Online Library}
}

@incollection{elwert2013graphical,
  title     = {Graphical causal models},
  author    = {Elwert, Felix},
  booktitle = {Handbook of causal analysis for social research},
  pages     = {245--273},
  year      = {2013},
  publisher = {Springer}
}

@book{brooks2011handbook,
  title     = {Handbook of {M}arkov {C}hain {M}onte {C}arlo},
  author    = {Brooks, Steve and Gelman, Andrew and Jones, Galin and Meng, Xiao-Li},
  year      = {2011},
  publisher = {CRC press}
}

@article{paolino2001maximum,
  title     = {Maximum likelihood estimation of models with beta-distributed dependent variables},
  author    = {Paolino, Philip},
  journal   = {Political Analysis},
  volume    = {9},
  number    = {4},
  pages     = {325--346},
  year      = {2001},
  publisher = {Cambridge University Press}
}

@book{wohlin2012experimentation,
  title     = {Experimentation in software engineering},
  author    = {Wohlin, Claes and Runeson, Per and H{\"o}st, Martin and Ohlsson, Magnus C and Regnell, Bj{\"o}rn and Wessl{\'e}n, Anders and others},
  volume    = {236},
  year      = {2012},
  publisher = {Springer}
}

@article{frattini2025applying,
  title     = {Applying bayesian data analysis for causal inference about requirements quality: a controlled experiment},
  author    = {Frattini, Julian and Fucci, Davide and Torkar, Richard and Montgomery, Lloyd and Unterkalmsteiner, Michael and Fischbach, Jannik and Mendez, Daniel},
  journal   = {Empirical Software Engineering},
  volume    = {30},
  number    = {1},
  pages     = {29},
  year      = {2025},
  publisher = {Springer},
  doi       = {10.1007/s10664-024-10582-1}
}

@book{cleland_software_2012,
  title     = {Software and Systems Traceability},
  author    = {{Cleland-Huang}, Jane and Gotel, Orlena and Zisman, Andrea and others},
  year      = {2012},
  volume    = {2},
  publisher = {Springer},
  doi       = {10.1007/978-1-4471-2239-5}
}

@article{charalampidou_empirical_2021,
  title      = {Empirical Studies on Software Traceability: {{A}} Mapping Study},
  shorttitle = {Empirical Studies on Software Traceability},
  author     = {Charalampidou, Sofia and Ampatzoglou, Apostolos and Karountzos, Evangelos and Avgeriou, Paris},
  year       = {2021},
  month      = feb,
  journal    = {Journal of Software: Evolution and Process},
  volume     = {33},
  number     = {2},
  pages      = {e2294},
  publisher  = {John Wiley \& Sons, Ltd},
  issn       = {2047-7481},
  doi        = {10.1002/smr.2294},
  urldate    = {2025-03-03},
  langid     = {english}
}

@inproceedings{vogelsang_impact_2025,
  title      = {On the {{Impact}} of {{Requirements Smells}} in {{Prompts}}: {{The Case}} of {{Automated Traceability}}},
  shorttitle = {On the {{Impact}} of {{Requirements Smells}} in {{Prompts}}},
  author     = {Vogelsang, Andreas and Korn, Alexander and Broccia, Giovanna and Ferrari, Alessio and Fischbach, Jannik and Arora, Chetan},
  year       = {2025},
  booktitle  = {Proceedings of the 2025 ACM/IEEE 45th International Conference on Software Engineering: New Ideas and Emerging Results}
}

@article{femmer_rapid_2017,
  title    = {Rapid Quality Assurance with {{Requirements Smells}}},
  author   = {Femmer, Henning and M{\'e}ndez Fern{\'a}ndez, Daniel and Wagner, Stefan and Eder, Sebastian},
  year     = {2017},
  month    = jan,
  journal  = {Journal of Systems and Software},
  volume   = {123},
  pages    = {190--213},
  issn     = {0164-1212},
  doi      = {10.1016/j.jss.2016.02.047},
  urldate  = {2025-03-05}
}

@inproceedings{femmer_which_2017,
  title      = {Which {{Requirements Artifact Quality Defects}} Are {{Automatically Detectable}}? {{A Case Study}}},
  shorttitle = {Which {{Requirements Artifact Quality Defects}} Are {{Automatically Detectable}}?},
  booktitle  = {2017 {{IEEE}} 25th {{International Requirements Engineering Conference Workshops}} ({{REW}})},
  author     = {Femmer, Henning and Unterkalmsteiner, Michael and Gorschek, Tony},
  year       = {2017},
  month      = sep,
  pages      = {400--406},
  doi        = {10.1109/REW.2017.18},
  urldate    = {2025-03-05}
}

@article{ferrari_detecting_2018,
  title      = {Detecting Requirements Defects with {{NLP}} Patterns: An Industrial Experience in the Railway Domain},
  shorttitle = {Detecting Requirements Defects with {{NLP}} Patterns},
  author     = {Ferrari, Alessio and Gori, Gloria and Rosadini, Benedetta and Trotta, Iacopo and Bacherini, Stefano and Fantechi, Alessandro and Gnesi, Stefania},
  year       = {2018},
  month      = dec,
  journal    = {Empirical Softw. Engg.},
  volume     = {23},
  number     = {6},
  pages      = {3684--3733},
  issn       = {1382-3256},
  doi        = {10.1007/s10664-018-9596-7},
  urldate    = {2025-03-05}
}

@inproceedings{hasso_detection_2019,
  title     = {Detection of Defective Requirements Using Rule-Based Scripts},
  booktitle = {International {{Conference}} on {{Requirements Engineering}} - {{Foundation}} for {{Software Quality}} ({{REFSQ}}) 2019},
  author    = {Hasso, H. and Geppert, H. and Dembach, M. and Toews, D.},
  year      = {2019},
  urldate   = {2025-03-05},
  langid    = {english}
}

@article{lami_automatic_2004,
  title    = {An {{Automatic Tool}} for the {{Analysis}} of {{Natural Language Requirements}}},
  author   = {Lami, Giuseppe and Gnesi, Stefania and Fabbrini, Fabrizio and Fusani, Mario and Trentanni, Gianluca},
  year     = {2004},
  langid   = {english}
}

@article{parra_methodology_2015,
  title    = {A Methodology for the Classification of Quality of Requirements Using Machine Learning Techniques},
  author   = {Parra, Eugenio and Dimou, Christos and Llorens, Juan and Moreno, Valent{\'i}n and Fraga, Anabel},
  year     = {2015},
  month    = nov,
  journal  = {Inf. Softw. Technol.},
  volume   = {67},
  number   = {C},
  pages    = {180--195},
  issn     = {0950-5849},
  doi      = {10.1016/j.infsof.2015.07.006},
  urldate  = {2025-03-05}
}

@article{rago_approach_2014,
  title     = {An {{Approach}} for {{Automating Use Case Refactoring}}},
  author    = {Rago, Alejandro and Frade, Paula and Ruiva, Miguel and Marcos, Claudia A.},
  year      = {2014},
  month     = jun,
  journal   = {Electronic Journal of SADIO},
  volume    = {vol. 13},
  issn      = {1514-6774},
  urldate   = {2025-03-05},
  copyright = {http://creativecommons.org/licenses/by/4.0/},
  langid    = {english}
}

@inproceedings{usdadiya_empirical_2019,
  title     = {An {{Empirical Study}} on {{Assessing}} the {{Quality}} of {{Use Case Metrics}}},
  booktitle = {Proceedings of the 12th {{Innovations}} in {{Software Engineering Conference}} (Formerly Known as {{India Software Engineering Conference}})},
  author    = {Usdadiya, Chirag and Tiwari, Saurabh and Banerjee, Asim},
  year      = {2019},
  month     = feb,
  series    = {{{ISEC}} '19},
  pages     = {1--11},
  publisher = {Association for Computing Machinery},
  address   = {New York, NY, USA},
  doi       = {10.1145/3299771.3299775},
  urldate   = {2025-03-05},
  isbn      = {978-1-4503-6215-3}
}

@article{ospina2012general,
  title={A general class of zero-or-one inflated beta regression models},
  author={Ospina, Raydonal and Ferrari, Silvia LP},
  journal={Computational Statistics \& Data Analysis},
  volume={56},
  number={6},
  pages={1609--1623},
  year={2012},
  publisher={Elsevier}
}

@article{mize2019general,
  title={A general framework for comparing predictions and marginal effects across models},
  author={Mize, Trenton D and Doan, Long and Long, J Scott},
  journal={Sociological Methodology},
  volume={49},
  number={1},
  pages={152--189},
  year={2019},
  publisher={SAGE Publications Sage CA: Los Angeles, CA}
}

@article{denters1989conditional,
  title={Conditional regression analysis: Problems, solutions and an application},
  author={Denters, Bas and Van Puijenbroek, Rob AG},
  journal={Quality and Quantity},
  volume={23},
  number={1},
  pages={83--108},
  year={1989},
  publisher={Springer}
}

@inproceedings{frattini2025adopting,
  title={Adopting Use Case Descriptions for Requirements Specification: an Industrial Case Study},
  author={Frattini, Julian and Frattini, Anja},
  booktitle={2025 IEEE 31st International Requirements Engineering Conference (RE)},
  year={2025},
  organization={IEEE}
}

@article{fernandez2012field,
  title={Field study on requirements engineering: Investigation of artefacts, project parameters, and execution strategies},
  author={Fernandez, Daniel Mendez and Wagner, Stefan and Lochmann, Klaus and Baumann, Andrea and de Carne, Holger},
  journal={Information and Software Technology},
  volume={54},
  number={2},
  pages={162--178},
  year={2012},
  publisher={Elsevier},
  doi={10.1016/j.infsof.2011.09.001}
}

@misc{gelman2018interaction,
  title={You need 16 times the sample size to estimate an interaction than to estimate a main effect},
  author={Gelman, Andrew},
  howpublished={\url{https://statmodeling.stat.columbia.edu/2018/03/15/need16/}},
  note={Accessed: 2023-11-24}
}

@inproceedings{femmer2014impact,
  title={On the impact of passive voice requirements on domain modelling},
  author={Femmer, Henning and Ku{\v{c}}era, Jan and Vetr{\`o}, Antonio},
  booktitle={Proceedings of the 8th ACM/IEEE international symposium on empirical software engineering and measurement},
  pages={1--4},
  year={2014}
}

@misc{replication-package,
  title = {{How Requirements Quality Makes (or Breaks) Traceability Link Recovery - Replication Package}},
  author={Hey, Tobias and Frattini, Julian},
  year={2026},
  howpublished = {\url{https://doi.org/10.5281/zenodo.20448214}},
}

@article{cohen1960coefficient,
  title={A coefficient of agreement for nominal scales},
  author={Cohen, Jacob},
  journal={Educational and psychological measurement},
  volume={20},
  number={1},
  pages={37--46},
  year={1960},
  publisher={Sage Publications Sage CA: Thousand Oaks, CA}
}

\end{document}